\documentclass[aps,prl,reprint,superscriptaddress,footinbib]{revtex4-2}

\usepackage{xcolor}
\usepackage{graphicx}
\usepackage{siunitx}
\usepackage{braket}
\usepackage{blkarray}
\usepackage{amsmath}
\usepackage{amsfonts}
\usepackage[normalem]{ulem}

\graphicspath{{Figures/}}

\begin{document}
\title{Entanglement of trapped-ion qubits separated by 230 meters} 

\author{V.~Krutyanskiy}
\affiliation{Institut f\"ur Quantenoptik und Quanteninformation, Osterreichische Akademie der Wissenschaften, Technikerstr. 21a, 6020 Innsbruck, Austria}
\affiliation{Institut f\"ur Experimentalphysik, Universit\"at Innsbruck, Technikerstr. 25, 6020 Innsbruck, Austria}

\author{M.~Galli}
\affiliation{Institut f\"ur Experimentalphysik, Universit\"at Innsbruck, Technikerstr. 25, 6020 Innsbruck, Austria}

\author{V.~Krcmarsky}
\affiliation{Institut f\"ur Quantenoptik und Quanteninformation, Osterreichische Akademie der Wissenschaften, Technikerstr. 21a, 6020 Innsbruck, Austria}
\affiliation{Institut f\"ur Experimentalphysik, Universit\"at Innsbruck, Technikerstr. 25, 6020 Innsbruck, Austria}

\author{S.~Baier}
\affiliation{Institut f\"ur Experimentalphysik, Universit\"at Innsbruck, Technikerstr. 25, 6020 Innsbruck, Austria}

\author{D.~A.~Fioretto}
\affiliation{Institut f\"ur Experimentalphysik, Universit\"at Innsbruck, Technikerstr. 25, 6020 Innsbruck, Austria}

\author{Y.~Pu}
\affiliation{Institut f\"ur Experimentalphysik, Universit\"at Innsbruck, Technikerstr. 25, 6020 Innsbruck, Austria}

\author{A.~Mazloom}
\affiliation{Department of Physics, Georgetown University, 37th and O Sts. NW, Washington, DC 20057, USA}

\author{P.~Sekatski}
\affiliation{Department of Applied Physics, University of Geneva, 1211 Geneva, Switzerland}

\author{M.~Canteri}
\affiliation{Institut f\"ur Quantenoptik und Quanteninformation, Osterreichische Akademie der Wissenschaften, Technikerstr. 21a, 6020 Innsbruck, Austria}
\affiliation{Institut f\"ur Experimentalphysik, Universit\"at Innsbruck, Technikerstr. 25, 6020 Innsbruck, Austria}

\author{M.~Teller}
\affiliation{Institut f\"ur Experimentalphysik, Universit\"at Innsbruck, Technikerstr. 25, 6020 Innsbruck, Austria}

\author{J.~Schupp}
\affiliation{Institut f\"ur Quantenoptik und Quanteninformation, Osterreichische Akademie der Wissenschaften, Technikerstr. 21a, 6020 Innsbruck, Austria}
\affiliation{Institut f\"ur Experimentalphysik, Universit\"at Innsbruck, Technikerstr. 25, 6020 Innsbruck, Austria}

\author{J.~Bate}
\affiliation{Institut f\"ur Experimentalphysik, Universit\"at Innsbruck, Technikerstr. 25, 6020 Innsbruck, Austria}

\author{M.~Meraner}
\affiliation{Institut f\"ur Quantenoptik und Quanteninformation, Osterreichische Akademie der Wissenschaften, Technikerstr. 21a, 6020 Innsbruck, Austria}
\affiliation{Institut f\"ur Experimentalphysik, Universit\"at Innsbruck, Technikerstr. 25, 6020 Innsbruck, Austria}

\author{N.~Sangouard}
\affiliation{Institut de Physique Th\'eorique, Universit\'e Paris-Saclay, CEA, CNRS, 91191 Gif-sur-Yvette, France}

\author{B.~P.~Lanyon}
\email[Correspondence should be send to]{ ben.lanyon@uibk.ac.at}
\affiliation{Institut f\"ur Quantenoptik und Quanteninformation, Osterreichische Akademie der Wissenschaften, Technikerstr. 21a, 6020 Innsbruck, Austria}
\affiliation{Institut f\"ur Experimentalphysik, Universit\"at Innsbruck, Technikerstr. 25, 6020 Innsbruck, Austria}

\author{T.~E.~Northup}
\affiliation{Institut f\"ur Experimentalphysik, Universit\"at Innsbruck, Technikerstr. 25, 6020 Innsbruck, Austria}

\date{\today}

\begin{abstract}

	We report on an elementary quantum network of two atomic ions separated by \SI{230}{\meter}. The ions are trapped in different buildings and connected with \SI{520(2)}{\meter} of optical fiber. At each network node, the electronic state of an ion is entangled with the polarization state of a single cavity photon; subsequent to interference of the photons at a beamsplitter, photon detection heralds entanglement between the two ions. 
Fidelities of up to $(88.2+2.3-6.0)\%$ are achieved with respect to a maximally entangled Bell state, with a success probability of $\SI{4e-5}{}$. 
We analyze the routes to improve these metrics, paving the way for long-distance networks of entangled quantum processors.

\end{abstract}
\maketitle

The realization of quantum networks~\cite{Kimble2008a, Wehner2018} that link cities and countries  would open up powerful new applications in information security~\cite{Pirandola2020}, 
distributed computing~\cite{Jiang07, Monroe2014},
precision sensing~\cite{Proctor2018, Sekatski2020} 
and timekeeping~\cite{Komar2014}. 
These applications require distributed quantum network nodes that, first, can be entangled via the exchange of photons over long distances and, second, can store and process quantum information encoded in registers of qubits.  
A handful of experiments have demonstrated remote entanglement of two quantum-logic-capable qubits, including ions in linear Paul traps~\cite{Moehring2007, Stephenson2020}, optically trapped neutral atoms~\cite{Ritter12, Hofmann12}, color centers in diamond~\cite{Bernien13}, quantum dots~\cite{Delteil2016, Stockill2017} and superconducting qubits~\cite{Magnard2020}; furthermore, three-node entanglement of color centers was recently achieved~\cite{Pompili2021}.  
These elementary networks have been extended to entangle quantum systems in separate buildings: two diamond color centers $\SI{1.3}{\kilo\meter}$ apart~\cite{Hensen2015} and two neutral atoms $\SI{400}{\meter}$ apart~\cite{Zhang2022, vanLeent2022}.

Quantum network nodes based on trapped ions \cite{Duan2010} promise high-fidelity quantum-gate operations on registers of tens of qubits~\cite{Bruzewicz2019, Friis2018}, coherence times exceeding one hour~\cite{WangYe2017}, efficient interfacing with telecom-wavelength photons~\cite{Bock2018, Krutyanskiy2019} and precision sensing and metrology~\cite{Brewer2019, Gilmore2021, Marciniak2022}.  
Building on the first demonstration of remote-ion entanglement~\cite{Moehring2007}, significant improvements in both rate and fidelity~\cite{Hucul2015, Stephenson2020} have recently enabled device-independent quantum key distribution~\cite{Nadlinger2021} and enhanced timekeeping~\cite{Nichol2021}, and a multispecies node has been demonstrated~\cite{Inlek2017}. 
Remote entanglement of trapped ions more than a few meters apart has not previously been reported. 

In this Letter, we report on the entanglement of two trapped ions separated by \SI{230}{\meter}. The two ions are in separate buildings, connected via \SI{520(2)}{\meter} of optical fiber, and controlled by independent lasers and electronics. Their entanglement is heralded by the coincident detection of two infrared photons that travel through the fiber. 
In contrast to implementations based on spontaneous emission~\cite{Moehring2007, Hucul2015, Inlek2017, Stephenson2020, Bernien13, Delteil2016, Stockill2017, Hofmann12, Zhang2022, vanLeent2022,Pompili2021},
our photon generation method is based on a cavity-mediated Raman process providing tunable entangled states \cite{Stute2012} and high efficiency~\cite{Schupp2021}, which are advantageous  
for establishing long-distance entanglement~\cite{Sangouard2009}. 
Remote ion--ion entanglement is characterized by quantum state tomography and analyzed for a range of time windows for coincident detection. 
A detailed model is developed that captures the observed trade-off between the fidelity of remote entanglement and the heralding efficiency and shows how significant improvements can be made in the future. 

\begin{figure}
	\includegraphics[width=0.47\textwidth]{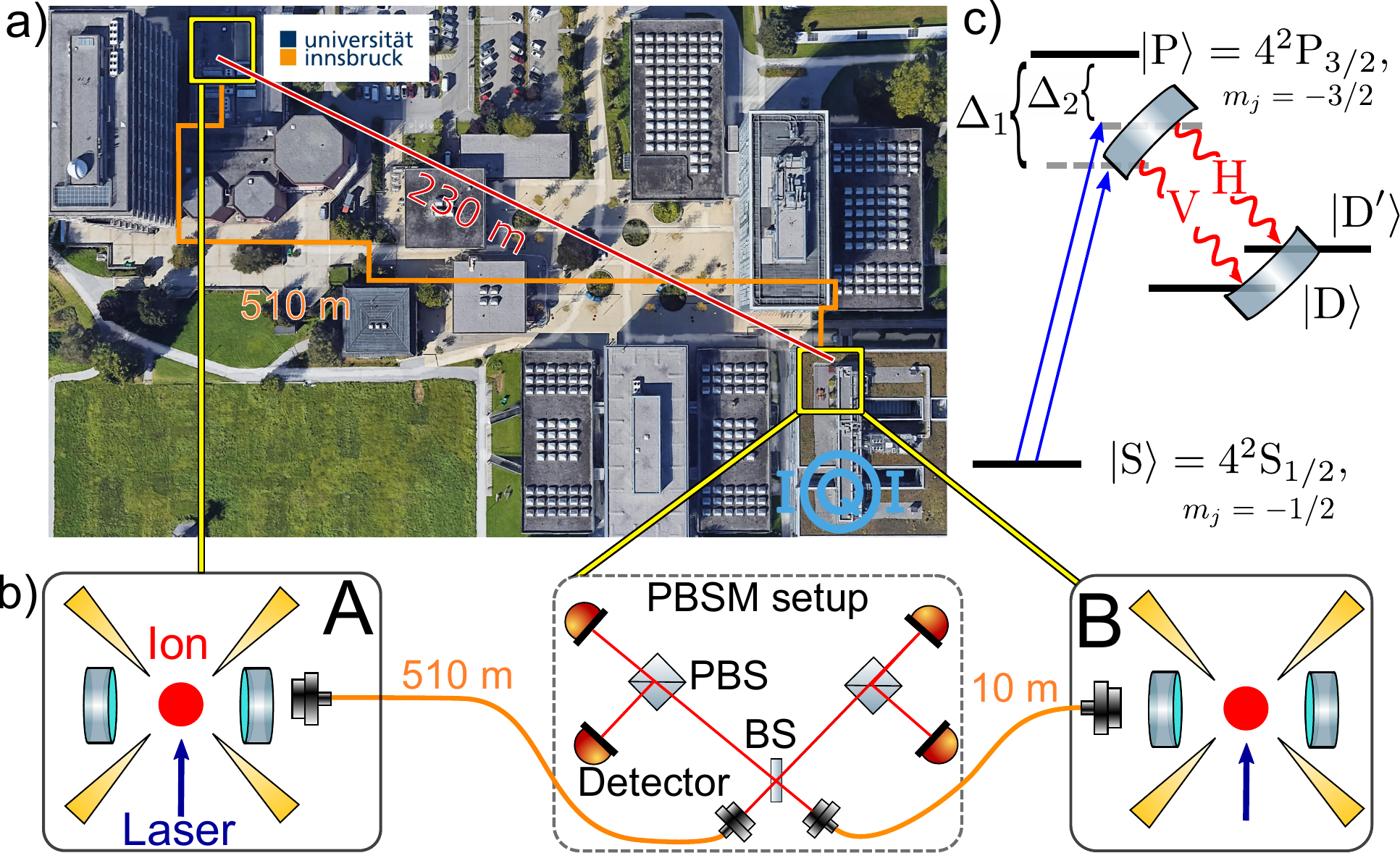}
	\caption{The two-node quantum network. a) Satellite image (Google Earth, image: Landsat / Copernicus). Nodes A and B are located in separate buildings, connected via a \SI{520(2)}{\meter} optical-fiber link and have a \SI{230}{\meter} line of sight separation. 
		b) 
		Nodes consist of an ion, a linear Paul trap (four yellow electrodes) and a cavity comprised of two mirrors. 
		The photonic Bell-state measurement (PBSM) setup contains a beam splitter (BS), polarizing beam splitters (PBS) and photon detectors.
		c) Energy-level diagram for $^{40}\text{Ca}^+$. 
		When an ion is in state $|S\rangle$ and no photons are in the cavity,
		a laser pulse containing two tones generates the ion--photon entangled state $1/\sqrt{2}\left (|\text{DV}\rangle  + e^{\mathrm{i} \theta} |\text{D}'\text{H}\rangle\right)$, where $|\text{V}\rangle$ and $|\text{H}\rangle$ are the polarization components of a cavity photon and $\theta$ is a phase \cite{Stute2012}. The frequency difference  $\Delta_2-\Delta_1$ is equal to the one between $|\text{D}'\rangle = 3^2\text{D}_{5/2}, m_j = -3/2$ and $|\text{D}\rangle = 3^2\text{D}_{5/2}, m_j = -5/2$. 
	}
	\label{fig:Fig.1}
\end{figure}

Each node in our quantum network (Fig.~\ref{fig:Fig.1}(a),(b)) consists of a single $^{40}\text{Ca}^+$ atom confined in a linear Paul trap and coupled to a $\SI{20}{\milli\meter}$ cavity for photon collection at $\SI{854}{\nano\meter}$. 
A photon is generated at each node via a bichromatic cavity-mediated Raman transition (Fig.~\ref{fig:Fig.1}(c))~\cite{Stute2012}. 
Here, a Raman laser pulse applied to the ion ideally generates the maximally entangled ion--photon state $\ket{\psi_{k}}=1/\sqrt{2}\left (|\text{D}\text{V}\rangle  + e^{\mathrm{i} \theta_k} |\text{D}'\text{H}\rangle\right)$, where $|\text{D}\rangle$ and $|\text{D}'\rangle$ are the respective Zeeman states $|3^2\text{D}_{5/2}, m_j = -5/2\rangle$ and $|3^2\text{D}_{5/2}, m_j = -3/2\rangle$, $|\text{V}\rangle$ and $|\text{H}\rangle$ are the vertical and horizontal polarization components of a photon emitted into the cavity vacuum mode, and $\theta_k$ is a phase set at node $k \in \{\text{A},\text{B}\}$. The photon exits the cavity and is coupled into single-mode optical fiber.
Two photons, one from each node, arrive at a photonic Bell-state measurement (PBSM) setup, where their spatial modes are overlapped on a balanced beamsplitter~\cite{Duan03,Feng03,Simon03}.
Coincident detection of orthogonally polarized photons ideally heralds the maximally entangled ion--ion states
\begin{equation} 
	|\Psi^{\pm} \rangle = 1/\sqrt{2} \left( \ket{\rm D_A D'_B}  \pm e^{\mathrm{i} \phi} \ket{\rm D'_A D_B} \right),
	\label{eq:eq1}
\end{equation}
with phase $\phi = \theta_\text{A} - \theta_\text{B}$, where subscripts indicate the ion node.
The state $|\Psi^+\rangle$ is obtained if the two coincident detection events occur in the same output mode of the beamsplitter, while $|\Psi^-\rangle$ is obtained if coincident detection occurs in opposite output modes.

Which-path information for the two photons should be erased in the PBSM, which requires both temporal and spectral indistinguishability of the photon wavepackets.
Each node has control software and hardware that executes a finite-length and node-specific remote entanglement sequence:  a list of operations to perform. 
Each control system is referenced to its own $\SI{10}{\mega\hertz}$ GPS clock source.
Temporal synchronization of the two sequences to within a $\SI{30}{\nano\second}$ jitter is achieved via a handshake between the control systems at the start of each sequence. 
The handshake signal is sent over a dedicated optical fiber in a fiber bundle connecting the two labs, which also contains the fiber for single-photon distribution.
Offsets in the arrival times of temporal photon wavepackets at the PBSM, e.g., due to optical path differences, are compensated for by introducing sequence delays. 

Spectral indistinguishability of the photons requires matching the resonant frequencies of the remote cavities.
This is achieved via periodic calibration at $\SI{20}{\minute}$ intervals:
$\SI{854}{\nano\meter}$ laser light that is resonant with the cavity at Node~A is sent to Node~B over a third fiber in the bundle, and the length of the Node~B cavity is adjusted until it is resonant with this light.
Also at $\SI{20}{\minute}$ intervals, the polarization rotation of the fiber that carries single photons is characterized and corrected for \footnote{See Supplemental Material at [URL will be inserted by publisher] for details on the synchronization and stabilization procedures between the two nodes, a full description of our density-matrix model, and the calculation of interference visibility, including Refs.~\cite{Russo2009,Stute2012a,Friebe2019,Schupp2021a,Casabone2015a,Casabone2013,Haeffner2008,Efron93,Craddock2019,Briegel2000,Horodecki99}.}.

\begin{figure}[t]
	\includegraphics[width=1\columnwidth]{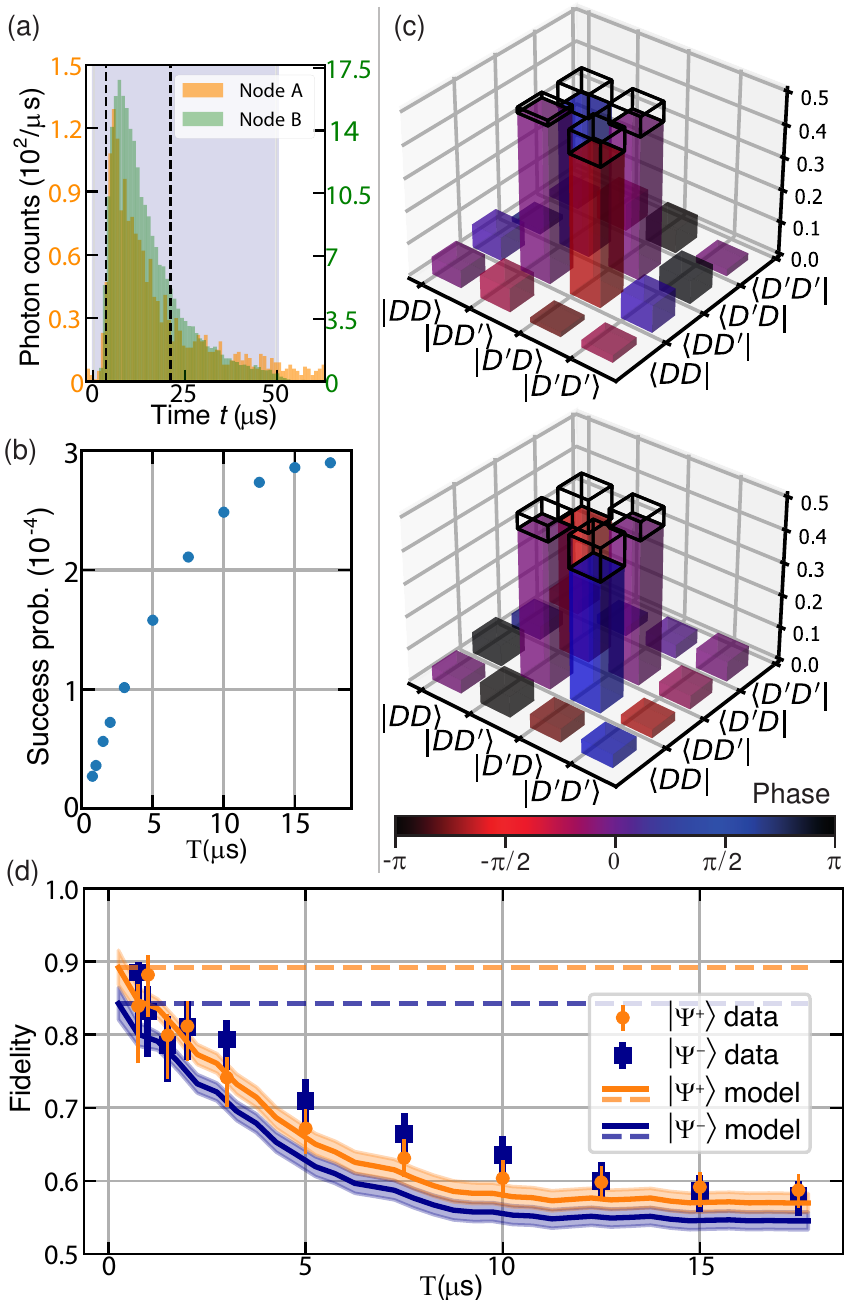}
	\caption{Entanglement between ion qubits. 
	(a) Single-photon wavepackets measured at each node in a separate calibration experiment.  Shown are histograms of photon counts per \SI{1}{\micro\second} time bin for ion-entangled photons from Node A only (orange) and Node B only (green). The gray region indicates when the Raman laser pulse is on.  The dashed black lines indicate the window within which  coincidence events are evaluated during entanglement experiments.
	(b) Success probability for a coincidence event heralding either $|\psi^+\rangle$ or $|\psi^-\rangle$ to occur as a function of $T$. 
	(c) Experimentally reconstructed density matrices $\rho^+(T)$ and $\rho^-(T)$, for  $T=\SI{1}{\micro\second}$. Bar heights indicate amplitudes of matrix entries; colors indicate phases.  Amplitudes of the entries for $| \Psi^{\pm}\rangle \langle \Psi^{\pm} |$ are outlined for comparison.
	(d) Fidelity $F^{\pm}$ as a function of $T$.  Markers indicate measured values; error bars correspond to one standard deviation.  Solid lines show an empirical 
	model discussed in the main text, with shaded regions indicating uncertainties. Dashed lines show a partial model omitting photon distinguishability.  
}
		\label{fig:Fig.2}
\end{figure}

The remote entanglement sequences at each node contain a loop in which up to 20 attempts are made to establish ion--ion entanglement.  
Each attempt contains \SI{0.3}{\milli\second} of state initialization, via Doppler cooling and optical pumping, followed by a Raman laser pulse of \SI{50}{\micro\second} to generate a photon. 
In the case of coincident photon detection within a \SI{50}{\micro\second} window that encompasses the single-photon wavepackets, the sequence exits the loop, and the ion qubits are measured.
Ion-qubit measurement consists of laser-driven single-qubit rotations to set the measurement basis, followed by state detection via electron shelving for \SI{1.5}{\milli\second}, at which point the sequence is concluded.

The remote ion--ion state is characterized via quantum state tomography, for which the sequence is repeated for all nine combinations of the Pauli measurement bases for two ion qubits~\cite{James2001}. 
Tomographic reconstruction, via the maximum likelihood technique, yields the density matrices $\rho^{\pm}(T)$, where $\rho^+$ and $\rho^-$ are reconstructed for the coincidences corresponding to ideally $|\Psi^+\rangle$ and $|\Psi^-\rangle$, respectively, and $T$ is the maximum time difference for which entanglement is heralded between coincident photons. 
A fidelity $F^{\pm}(T) \equiv \langle \Psi^{\pm}| \rho^{\pm}(T) | \Psi^{\pm} \rangle > 0.5$ proves entanglement of the remote ions.
Uncertainties for $F^{\pm}(T)$ and for all quantities derived from the density matrices are obtained via Monte Carlo resampling~\cite{Note1}. 

Data were acquired over seven hours, including interspersed calibrations. For each basis measurement setting, $\SI{17}{\min}$ of data were acquired on average.
In total, $\SI{13656928}{}$ attempts were made to generate remote entanglement, resulting in $\SI{3960}{}$
coincidence events within the interval $[t = \SI{5.5}{\micro\second},t = \SI{23}{\micro\second}]$~(Fig.~\ref{fig:Fig.2}(a)), corresponding to a $\SI{0.029}{\percent}$ probability of two-photon coincidence per attempt, which we define as the success probability.
Here $t=0$ indicates the start of the \SI{50}{\micro\second} detection window, and the narrower interval has been chosen to improve signal to noise.
The remote entanglement rate during the data acquisition time is thus $\SI{0.43}{\sec^{-1}}$.
The fidelities of the reconstructed states are $F^+(\SI{17.5}{\micro\second}) = (58.7 + 2.1 - 2.3)\%$ and $F^-(\SI{17.5}{\micro\second}) = (58.0 + 2.5 - 2.9)\%$, 
where $T=\SI{17.5}{\micro\second}$ corresponds to all possible coincidences within the \SI{17.5}{\micro\second} window.

When we take a subset of the data, corresponding to coincidences separated by smaller values of $T$, entangled ion--ion states are generated with higher fidelity, at the cost of a lower success probability (Fig.~\ref{fig:Fig.2}(b)).
The density matrices shown in Fig.~\ref{fig:Fig.2}(c) correspond to $T = \SI{1}{\micro\second}$, for which we recorded 491
coincidence events, that is, a remote entanglement rate of $\SI{3.1}{\min^{-1}}$. The fidelities of the reconstructed states are
$F^+(\SI{1}{\micro\second}) = (88.2+2.3-6.0)\%$ and $F^-(\SI{1}{\micro\second}) = (82.2+2.5-6.4)\%$. 
We optimize $F^{\pm}$ over the phase $\phi$ in Eq.~\ref{eq:eq1} because we did not determine $\theta_A$ and $\theta_B$ independently; this optimization yields $\phi = \SI{81.2}{\degree}$. 
We then fix this value of $\phi$ for all subsequent data points.
In Fig.~\ref{fig:Fig.2}(d), we plot the measured fidelities for values of $T$ between $\SI{0.75}{\micro\second}$ and $\SI{17.5}{\micro\second}$.

A decrease in fidelity as $T$ increases is to be expected:  for example, spontaneous emission during the Raman process provides information on which ion generated which cavity photon~\cite{Casabone2015, Walker2020, Meraner2020}, that is, scattering introduces which-path information.
To predict our experimentally determined fidelities, we have developed an empirical model for the ion--ion density matrix heralded by two-photon detection. The model contains photon distinguishability along with two other sources of infidelity: detector background counts and imperfect ion--photon entanglement.  The values of $F^{\pm}(T)$ calculated using this density-matrix model are plotted in Fig.~\ref{fig:Fig.2}(d) along with the measured values. We will first explain the contributions of photon distinguishability to this model and will afterwards discuss the other sources of infidelity~\cite{Note1}.

\begin{figure}
	\includegraphics[width=0.45\textwidth]{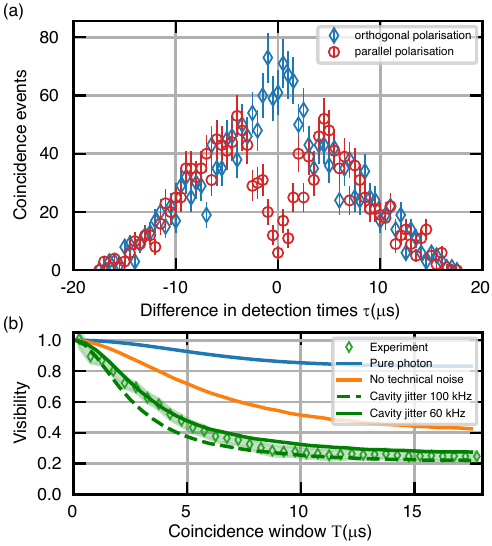}
	\caption{
		(a) Number of coincidences recorded for orthogonal (blue) and parallel (red) polarization projections of photons from Nodes A and B, for the same data set as in Fig.~\ref{fig:Fig.2}. Data are plotted as a function of the time difference $\tau$ between photon detection events, binned in \SI{0.5}{\micro\second}~intervals.  Error bars indicate Poissonian statistics.
	(b) Diamonds show the two-photon interference visibility calculated from the coincidence data after correction for background counts and detector efficiencies, using Eq.~\ref{eq:eq2}. The shaded region indicates the propagation of Poissonian uncertainties.  Lines show a master-equation model discussed in the main text.}
	\label{fig:Fig.3}
\end{figure}

To account for photon distinguishability, we employ a two-qubit dephasing channel, 
which reduces the off-diagonal elements of the ideal density matrices $| \Psi^{\pm}\rangle \langle \Psi^{\pm} |$~\cite{Note1}.
 The probability for dephasing in the channel is parameterized by the Hong-Ou-Mandel (HOM) interference visibility, which provides direct information about photon indistinguishability~\cite{Trivedi2020}.  For unit visibility, no dephasing occurs, while for a visibility of zero,  all off-diagonal matrix elements are zero.

The HOM visibility is extracted from the photon coincidence data by sorting all events into two sets: coincidences with identical polarization and with orthogonal polarization.
Photons with identical polarization will exit the balanced beamsplitter at the same output port if they are otherwise indistinguishable, generating a HOM dip in coincidence counts at the two output ports~\cite{Hong1987}.
Orthogonally polarized photons are distinguishable and thus exhibit no HOM effect; their cross-correlation function allows us to normalize the HOM dip and thereby to calculate the interference visibility.
In Fig.~\ref{fig:Fig.3}(a), the number of coincidence events is plotted for both sets of data as a function of the time difference $\tau$ between photon detection events, for a time bin $\delta = \SI{0.5}{\micro\second}$.
The HOM dip at $\tau = 0$ can be clearly observed.
The interference visibility is then obtained from the data of Fig.~\ref{fig:Fig.3}(a) via the following procedure:
First, the number of expected coincidences between photons and detector background counts is subtracted from the number of measured coincidences for each time bin.  Next, the data sets are corrected for the detector efficiencies, which have been independently measured~\cite{Note1}.  
We define $N_{\parallel, k\delta}$ and $N_{\perp, k\delta}$ as the corrected numbers of coincidences for the time bin centered at $\tau = k\delta$ for $k \in \mathbb{Z}$, where the symbols $\parallel$ and $\perp$ indicate identical and orthogonal photon polarization, respectively.  Finally, the interference visibility is calculated as a function of the coincidence window $T$:
\begin{equation}
V(T) = 1-\frac{\sum_{(-T + \delta/2) \leq k\delta \leq (T - \delta/2)} 
	N_{\parallel,k\delta }
}
{\sum_{(-T + \delta/2) \leq k\delta \leq (T - \delta/2)}  
	N_{\perp,k\delta}
}
\label{eq:eq2}
\end{equation}
In Fig.~\ref{fig:Fig.3}(b), $V(T)$ is plotted for coincidence windows up to \SI{17.5}{\micro\second}, as in Fig.~\ref{fig:Fig.2}(d), and for $\SI{0.5}{\micro\second}$ time bins.  The maximum visibility corresponds to \SI{101(6)}{\percent} for $T=\SI{0.25}{\micro\second}$.

Our empirical model also includes detector background counts and imperfect ion--photon entanglement. 
For background counts, we use a white-noise channel based on the independently measured count rates of the four detectors. 
For ion--photon entanglement, we assume that imperfections   translate as a two-qubit depolarizing channel on the ion--ion state~\cite{Note1}.  
Ion--photon entanglement was characterized in a calibration measurement at each node via quantum state tomography immediately prior to ion--ion entanglement, and fidelities of $(92.9+0.4-0.5)\%$ and $(95.5+0.6-0.9)\%$ with respect to a maximally entangled state were obtained at Nodes A and B, respectively.

The empirical model is used to calculate the theoretical fidelities of Fig.~\ref{fig:Fig.2}(d): The solid lines are calculated from the full model, taking into account all three sources of infidelity, while the dashed lines are calculated when photon distinguishability is excluded from the model. Different values are predicted for $F^+$ and $F^-$ due to the use of superconducting nanowire detectors at two of the four beamsplitter outputs, which have lower dark-count rates than the single-photon-counting modules at the other two outputs. Based on the agreement between measured and modeled fidelities in Fig.~\ref{fig:Fig.2}(d), we conclude that the model captures the relevant properties of our setup and that the observed decline in fidelity as a function of $T$ is due to the corresponding decline in visibility.

For insight into how, in the future, visibility could be maintained for larger coincidence windows---thereby increasing the probability to establish remote entanglement with a given fidelity---we have developed a master-equation model based on that of Ref.~\cite{Meraner2020}.
This model considers three independently estimated noise processes that 
result in non-transform-limited (and therefore distinguishable) photons at each node:
frequency jitter of the Node A cavity by $\SIrange[range-phrase=-,range-units=single]{60}{100}{\kilo\hertz}$, Raman-laser phase noise, and spontaneous emission. 
All parameter values used in the model are statistically consistent with independent estimates and determined via comparison of the model to measured single-photon wavepackets~\cite{Note1}.
The predicted visibility is plotted in Fig.~\ref{fig:Fig.3}(b) for upper and lower estimates of the frequency jitter (green lines). It can be seen that the model is consistent with the visibility data.

We now look to the master-equation model to understand the impact of future improvements. 
Setting the technical noise contributions of cavity jitter and laser phase noise to zero, as shown in orange in Fig.~\ref{fig:Fig.3}(b), improves the model visibility. 
In addition, selecting only those ion--photon entanglement events for which no spontaneous emission occurs, corresponding to transform-limited photons,
leads to the most significant improvement in the model visibility (blue line). 
The remaining visibility imperfections are due to mismatch between the temporal wavepackets of the transform-limited photons 
produced at each node.

With regards to the technical noise contributions, we expect to suppress both cavity jitter and laser phase noise to negligible levels by improving the lock electronics and the passive cavity used as a laser reference at Node~A. Meanwhile, temporal wavepacket mismatch can be addressed through amplitude shaping of the Raman laser pulse~\cite{Fioretto2020}. 
It is spontaneous emission that poses the most significant challenge. Using our existing setup, multi-ion superradiant states can be harnessed to boost the fraction of photons generated without prior spontaneous decay~\cite{Casabone2015}; in future ion--cavity nodes, further gains can be obtained through judicious choice of the mirror properties and cavity geometry~\cite{Schupp2021}.
All these steps will increase the probability to generate transform-limited photons in each entanglement attempt. Additional steps can be taken to increase the attempt rate, namely, in the short term, implementing more efficient cooling and state detection protocols, and in the long term, coupling ions to fiber-based cavities with stronger coherent coupling and faster decay rates~\cite{Kobel2021}.
It is notable that the success probabilities shown in Fig.~\ref{fig:Fig.2}(b) are comparable to those achieved over a few meters in Ref.~\cite{Stephenson2020}, and that in future long-distance networks limited by photon travel time, it will be success probabilities that determine entanglement rates~\cite{Sangouard2009}.

In conclusion, we have verified entanglement over the longest trapped-ion network to date, with fidelities up to $(88.2+2.3-6.0)\%$ with respect to a maximally entangled state. A trade-off between fidelity and coincidence-window length was explained with the help of two models: an empirical model for the two-ion density matrix and a master-equation model to predict the interference visibility. Based on these models, we anticipate that we will be able to obtain significantly higher rates across this cavity-mediated network while maintaining high fidelities. 
Furthermore, in contrast to prior remote-ion entanglement~\cite{Moehring2007,Hucul2015,Stephenson2020}, which has been mediated by ultraviolet photons, our use of infrared photons enables efficient single-stage quantum frequency conversion to the telecom C band~\cite{Krutyanskiy2019}, providing a direct route to extend the quantum channel to hundreds of kilometers.
While the experiments presented here relied on just one ion at each node, a particular strength of the trapped-ion platform is the capability for quantum-information processing with dozens of addressed qubits in a single trap~\cite{Bruzewicz2019, Friis2018} and fidelities sufficient for fault-tolerant gate operations and error correction~\cite{RyanAnderson2021, Postler2022}.
This capability provides a route to robust logical qubit encodings at network nodes~\cite{Zwerger2017}, separate communication and information processing functionalities within each node~\cite{Duan2010,Monroe2014}, and quantum repeaters requiring Bell state measurements and either purification or error correction~\cite{Munro2015}.\\

\begin{acknowledgments}
This work was supported by
the European Union’s Horizon 2020 research and innovation program under grant agreement No. 820445 and project name ``Quantum Internet Alliance,"
by Projects F 7109 and Y 849 of the Austrian Science Fund (FWF),
and by the U.S. Army Research Laboratory's Center for Distributed Quantum Information under Cooperative Agreement Number W911NF-15-2-0060.
We acknowledge funding for S.B. by an FWF Erwin Schr\"odinger fellowship (No.~J 4229),
for V. Krutyanskiy by the Erwin Schrödinger Center for Quantum Science \& Technology (ESQ) Discovery Programme,
for B.P.L. by the CIFAR Quantum Information Science Program of Canada,
for A.M. by the U.S. National Science Foundation under Grant No. PHY-1915130,
for N.S. by the Commissariat \`a l'Energie Atomique et aux Energies Alternatives (CEA),
and for M.T. by the Early Stage Funding Program of the Vice-Rectorate for Research of the University of Innsbruck.

M.G. and V.~Krutyanskiy contributed equally to this work.
S.B., D.F., M.G., V.~Krcmarsky, V.~Krutyanskiy, M.M., Y.P., and J.S.
contributed to the experimental setup. 
S.B., J.B., M.C., D.F., M.G., V.~Krcmarsky, V.~Krutyanskiy, Y.P., and M.T. took data.
S.B., M.C., M.G., and V.~Krutyanskiy  
analyzed data.	
S.B., J.B., D.F.,
V.~Krcmarsky, V.~Krutyanskiy, A.M., Y.P., N.S., J.S., and P.S. 
performed theoretical modeling. 
B.P.L. and T.E.N. wrote the main text, with
contributions from all authors. 
The project was conceived and
supervised by B.P.L. and T.E.N. \\
\end{acknowledgments}

\bibliography{bibliography}

\end{document}


\title{Entanglement of trapped-ion qubits separated by 230 meters: Supplemental material}

\author{V.~Krutyanskiy}
\affiliation{Institut f\"ur Quantenoptik und Quanteninformation, Osterreichische Akademie der Wissenschaften, Technikerstr. 21a, 6020 Innsbruck, Austria}
\affiliation{Institut f\"ur Experimentalphysik, Universit\"at Innsbruck, Technikerstr. 25, 6020 Innsbruck, Austria}

\author{M.~Galli}
\affiliation{Institut f\"ur Experimentalphysik, Universit\"at Innsbruck, Technikerstr. 25, 6020 Innsbruck, Austria}

\author{V.~Krcmarsky}
\affiliation{Institut f\"ur Quantenoptik und Quanteninformation, Osterreichische Akademie der Wissenschaften, Technikerstr. 21a, 6020 Innsbruck, Austria}
\affiliation{Institut f\"ur Experimentalphysik, Universit\"at Innsbruck, Technikerstr. 25, 6020 Innsbruck, Austria}

\author{S.~Baier}
\affiliation{Institut f\"ur Experimentalphysik, Universit\"at Innsbruck, Technikerstr. 25, 6020 Innsbruck, Austria}

\author{D.~A.~Fioretto}
\affiliation{Institut f\"ur Experimentalphysik, Universit\"at Innsbruck, Technikerstr. 25, 6020 Innsbruck, Austria}

\author{Y.~Pu}
\affiliation{Institut f\"ur Experimentalphysik, Universit\"at Innsbruck, Technikerstr. 25, 6020 Innsbruck, Austria}

\author{A.~Mazloom}
\affiliation{Department of Physics, Georgetown University, 37th and O Sts. NW, Washington, DC 20057, USA}

\author{P.~Sekatski}
\affiliation{Department of Applied Physics, University of Geneva, 1211 Geneva, Switzerland}

\author{M.~Canteri}
\affiliation{Institut f\"ur Quantenoptik und Quanteninformation, Osterreichische Akademie der Wissenschaften, Technikerstr. 21a, 6020 Innsbruck, Austria}
\affiliation{Institut f\"ur Experimentalphysik, Universit\"at Innsbruck, Technikerstr. 25, 6020 Innsbruck, Austria}

\author{M.~Teller}
\affiliation{Institut f\"ur Experimentalphysik, Universit\"at Innsbruck, Technikerstr. 25, 6020 Innsbruck, Austria}

\author{J.~Schupp}
\affiliation{Institut f\"ur Quantenoptik und Quanteninformation, Osterreichische Akademie der Wissenschaften, Technikerstr. 21a, 6020 Innsbruck, Austria}
\affiliation{Institut f\"ur Experimentalphysik, Universit\"at Innsbruck, Technikerstr. 25, 6020 Innsbruck, Austria}

\author{J.~Bate}
\affiliation{Institut f\"ur Experimentalphysik, Universit\"at Innsbruck, Technikerstr. 25, 6020 Innsbruck, Austria}

\author{M.~Meraner}
\affiliation{Institut f\"ur Quantenoptik und Quanteninformation, Osterreichische Akademie der Wissenschaften, Technikerstr. 21a, 6020 Innsbruck, Austria}
\affiliation{Institut f\"ur Experimentalphysik, Universit\"at Innsbruck, Technikerstr. 25, 6020 Innsbruck, Austria}

\author{N.~Sangouard}
\affiliation{Institut de Physique Th\'eorique, Universit\'e Paris-Saclay, CEA, CNRS, 91191 Gif-sur-Yvette, France}

\author{B.~P.~Lanyon}
\email[Correspondence should be send to]{ ben.lanyon@uibk.ac.at}
\affiliation{Institut f\"ur Quantenoptik und Quanteninformation, Osterreichische Akademie der Wissenschaften, Technikerstr. 21a, 6020 Innsbruck, Austria}
\affiliation{Institut f\"ur Experimentalphysik, Universit\"at Innsbruck, Technikerstr. 25, 6020 Innsbruck, Austria}

\author{T.~E.~Northup}
\affiliation{Institut f\"ur Experimentalphysik, Universit\"at Innsbruck, Technikerstr. 25, 6020 Innsbruck, Austria}

\renewcommand{\theequation}{S\arabic{equation}}

\date{\today}

\begin{abstract}
\end{abstract}

\maketitle

\section{Ion-trap network nodes}

\paragraph*{Overview.}
\label{sec:overview}
The ion-trap network nodes are both in room-temperature vacuum chambers and employ the same basic design. Specifically, a macroscopic linear Paul trap is rigidly suspended from the top flange of each vacuum chamber; thus, the ion's motional mode along the trap's axis of symmetry (the axial mode) is vertical, and the two other modes (radial modes) lie in the horizontal plane. An in-vacuum optical cavity around the ion trap is mounted via nanopositioning stages on the bottom flange of each chamber; the cavity axis is a few degrees off horizontal. Both cavities are \SI{20}{\milli\meter} long and in the near-concentric regime, corresponding to microscopic waists at the ion location. Ions are loaded into each trap using a resistively heated oven of atomic calcium and a two-photon ionization process driven by lasers at \SI{422}{\nano\meter} and \SI{375}{\nano\meter}. Details on Node A can be found in~\cite{Russo2009, Stute2012a, Friebe2019}. Details on Node B can be found in~\cite{Krutyanskiy2019, Schupp2021, Schupp2021a}.

\paragraph*{Cavity parameters.}
At Node A, the transmission of the cavity mirrors at 854 nm was measured to be \SI{13(1)}{ppm} for the output mirror and \SI{1.3(3)}{ppm} for the second mirror~\cite{Stute2012a}, with a probability of \SI{20(2)}{\percent} that a photon in the cavity mode leaves the cavity through the output mirror~\cite{Casabone2015a}. 
At Node B, the measured transmission values at \SI{854}{\nano\meter} are \SI{90(4)}{ppm} for the output mirror and \SI{2.9(4)}{ppm} for the second mirror, 
and the probability that a photon leaves through the output mode is \SI{78(3)}{\percent}~\cite{Schupp2021}. The decay rates of the cavity fields, measured via cavity ringdown, are $\kappa_A = 2\pi \times \SI{68.4(6)}{\kilo\hertz}$~\cite{Friebe2019} and $\kappa_B = 2\pi \times \SI{70(2)}{\kilo\hertz}$~\cite{Schupp2021}.

\paragraph*{Trap frequencies.} At Node A, the frequencies of the axial and radial modes are $(\omega_\text{ax},\omega_\text{r1},\omega_\text{r2}) = 2\pi \times (1.13,1.70,1.76)\,{\rm MHz}$.
At Node B, they are  $2\pi \times (0.92,2.40,2.44)\,\rm{MHz}$~\cite{Schupp2021}.

\paragraph*{Ion-cavity geometry.} 
For the remaining discussions in this section, we use a Cartesian coordinate system with three orthogonal axes: $x$, $y$ and $z$.
At each node, the $z$ axis is the ion trap's axis of symmetry, defined by the line connecting the trap's DC endcap electrodes, which is the axis of the ion's motion at frequency $\omega_\text{ax}$. 
The $xz$ plane is defined as the plane containing both the $z$ axis and the cavity axis. The cavity axis subtends an angle with respect to the $x$ axis of \SI{4}{\degree} at both Node~A and Node~B~\cite{Stute2012a,Casabone2013, Schupp2021a}.

\paragraph*{Quantization axis.}  At each node, the atomic quantization axis is chosen to be parallel to the axis of an applied static magnetic field. This magnetic-field axis is set to subtend an angle of \SI{45}{\degree} with respect to the $z$ axis and to be perpendicular to the cavity axis; at Node B, it is likely that it is a few degrees off from perpendicular. 
At Node A, a magnetic field of \SI{4.2303(2)}{G} is set by DC currents in coils attached to the outside of the vacuum chamber. At Node B, a magnetic field of \SI{4.1713(4)}{G} is set by permanent magnets attached to the outside of the vacuum chamber. Both field strengths are measured via Ramsey spectroscopy of a single ion.

\paragraph*{Laser beam geometry.} A bichromatic laser field at \SI{393}{\nano\meter} drives the cavity-mediated Raman transition. At each node, the propagation direction of the Raman laser field is parallel to the magnetic-field axis. The field is circularly polarized in order to maximize the coupling strength on the $\ket{S}\equiv\ket{4^2S_{1/2},m_j=-1/2}$ to $\ket{P}\equiv\ket{4^2P_{3/2},m_j=-3/2}$ transition. This coupling is depicted in Fig.~1c of the main text. 

At Node A, Doppler cooling and state detection are implemented using \SI{397}{\nano\meter} laser fields along two axes and a \SI{866}{\nano\meter} field along a third axis.  Optical pumping and ion-qubit rotations are implemented using a \SI{729}{\nano\meter} field that lies in the $xz$ plane at an angle of \SI{45}{\degree} with respect to the $z$ axis.  

At Node B, Doppler cooling is implemented using a single beam path that lies in the $xz$ plane at an angle of \SI{45}{\degree} with respect to the $z$ axis, along which both \SI{397}{\nano\meter} and \SI{866}{\nano\meter} laser fields are sent.  Optical pumping is implemented using a second, circularly polarised, \SI{397}{\nano\meter} laser field in a direction parallel to the magnetic-field axis.  Ion-qubit rotations are implemented using a \SI{729}{\nano\meter} field at an angle of \SI{45}{\degree} with respect to the $z$ axis.

\section{Fiber-optic channels}
\label{sec:channels}
\paragraph*{Fiber bundles.} The laboratories in which Nodes A and B are located are connected with two optical fiber bundles, each of which contains eight single-mode optical fibers.
The bundles are installed along the same path between the laboratories, which follows underground corridors but includes a section several tens of meters in length that is exposed to outdoor air. 
Three optical signals are sent between the laboratories using the fiber bundles, each in a different fiber:
\begin{enumerate}
	\item \SI{854}{\nano\meter} single photons, 
	\item \SI{1550}{\nano\meter} laser light carrying digital trigger signals, 
	\item \SI{854}{\nano\meter} laser light that is used to match the resonance frequencies of the cavities.
\end{enumerate} 
Signal~1 is sent through one of the bundles. Signals~2 and 3 are sent through different fibers in the other bundle. None of the fibers are polarization maintaining.

\paragraph*{Stabilization of fiber polarization dynamics.} Signal~1 consists of single photons that travel from Node A over one fiber bundle and through local fiber extensions to reach the photonic Bell-state measurement (PBSM) setup introduced in the main text. Every 20 minutes during attempts to generate remote ion entanglement, the polarization rotation of this fiber channel is characterized and corrected for, a process that takes a few minutes. 

The polarization rotation is characterized via quantum process tomography, for which six input states are injected sequentially into the channel: single photons with horizontal, vertical, diagonal, antidiagonal, right-circular, and left-circular polarizations. The single photons are produced at Node A via a monochromatic cavity-mediated Raman process that is repumped continuously at \SI{854}{\nano\meter}; this process generates linearly polarized photons with a measured contrast ratio of 10.5:1.
After exiting the vacuum chamber, the photons pass through motorized waveplates, which we use to prepare the six input states.  

For each input state, the output state is analyzed using existing components at the PBSM setup (a polarizing beam splitter and photon detectors) along with additional waveplates. We perform measurements in sufficiently many bases to reconstruct each output state via quantum state tomography.  
A numerical search is then carried out over the data from all six states to find the nearest unitary polarization rotation, which we identify as the transformation of the fiber channel. Finally, at the input to the PBSM setup, the angles of three waveplates---a half-waveplate sandwiched by two quarter-waveplates---are set so that collectively, the waveplates implement the inverse of the unitary operation, thereby correcting for the transformation of the channel.  

\section{Photonic Bell-State Measurement (PBSM) setup}
\label{sec:PBSM}
A simplified schematic of the PBSM setup is shown in Fig.~1b of the main text.
The three waveplates described in the previous paragraph are not depicted in the figure. They are located between the output fiber coupler from Node A and the nonpolarizing beamsplitter. Two additional waveplates---also not depicted---are located between the output fiber coupler from Node B and the nonpolarizing beamsplitter. They consist of a quarter-waveplate and a half-waveplate and are used for calibration and analysis of the ion--photon state from Node A.

As shown in Fig.~1b, the PBSM setup has four single-photon detectors: two for each output mode of the nonpolarizing beamsplitter. 
In one of the beamsplitter output paths, the two detectors are single-photon counting modules (SPCMs); in the other output path, they are superconducting nanowire single-photon detectors (SNSPDs).

To determine the background counts and efficiency for each detector, we execute the same sequence as used for ion--ion entanglement (described in detail in Sec.~\ref{sec:exp_seq}) with one difference: photon detection does not terminate the photon-generation loop. In order to evaluate the values from each node separately, we  block the beam path from the other node. First, we define the background window as the interval $[t = \SI{70}{\micro\second},t = \SI{100}{\micro\second}]$, where, as in the main text, $t = 0$ indicates the start of the \SI{50}{\micro\second} detection
window.  No photons generated by an ion are expected in this window as the Raman laser pulse has been off for at least \SI{20}{\micro\second}. We determine the mean value of background counts per second as well as the probability of a background count during the detection window  $p_{{\rm bg-det}_r}$, where the detection window is defined as $[t = \SI{5.5}{\micro\second},t = \SI{23}{\micro\second}]$.

Next, we determine the mean photon number within the detection window and subtract $p_{\rm bg}$, yielding the probability $p^{k}_{{\rm det}_r}$ of detecting a photon at detector $r$ due to the Raman process at node $k \in \{\text{A},\text{B}\}$ within this  window. All values are summarized in Tab.~\ref{tab:detectors}. 
These values are used in the empirical model of Sec.~\ref{Empirical_model} in order to evaluate the influence of background counts on the ion-ion density matrices.
\newcolumntype{M}[1]{>{\centering\arraybackslash}m{#1}}
\begin{table}[h!]
	\begin{center}
		\begin{tabular}{|M{1.5cm}||M{1.7cm} M{1.6cm} M{1.4cm} M{1.4cm}||} 
			\hline \vspace{2ex}
			detector $r$ & \centering background (1/s) & $p_{{\rm bg-det}_r} ($\%$)$ & $p^{\rm A}_{{\rm det}_r} ($\%$)$ & $p^{\rm B}_{{\rm det}_r} ($\%$)$ \\ [1ex] 
			\hline\hline \vspace{2ex}
			SPCM$_1$ & 9.69 & 0.017 & 0.08 & 1.30 \\ [1ex] 
			\hline \vspace{2ex} 
			SPCM$_2$ & 9.37 & 0.016 & 0.12 & 1.96 \\ [1ex] 
			\hline \vspace{2ex} 
			SNSPD$_1$ & 0.25 & 0.0004 & 0.19 & 2.82 \\ [1ex] 
			\hline \vspace{2ex}
			SNSPD$_2$ & 2.00 & 0.0035 & 0.24 & 3.62 \\ [1ex] 
			\hline
		\end{tabular}
		\caption{Background counts, background-count probability within each detection window, and background-subtracted detection probability for each node, for each of the four detectors.}
		\label{tab:detectors}
	\end{center}
\end{table}

\section{Experimental sequences}
\label{sec:exp_seq}
\paragraph*{Initialization and handshake.}
At each node, we implement a finite-length and node-specific sequence.
The sequences at both Nodes A and B begin with Doppler cooling a single ion for at least \SI{1.52}{\milli\second}. 
Subsequently, 
\begin{enumerate}
	\item Node A sets TTL$^{A\rightarrow B}$ high on a \SI{1550}{\nano\meter} communication channel to Node B (Signal 2 in Sec.~\ref{sec:channels}).
	\item Upon receipt of the high TTL$^{A\rightarrow B}$, Node B sets TTL$^{B\rightarrow A}$ high on another communication channel on the same optical fiber to Node A. (The optical multiplexer supports four communication channels on one fiber.)
	\item Upon receipt of the high TTL$^{B\rightarrow A}$, Node A sets TTL$^{A\rightarrow B}$ to low.
	\item Upon receipt of the low TTL$^{A\rightarrow B}$, Node B sets TTL$^{B\rightarrow A}$ to low, completing the handshake. 
\end{enumerate}
Appropriate wait times are added between the operations to allow for processing and signal travel time at both nodes. 
The shortest time for a handshake is about \SI{10}{\micro\second}. 
We estimate remote clock-frequency mismatch of at most \SI{50}{\milli\hertz}, which has a negligible effect on sequence synchronization given the maximum sequence length of \SI{11.9}{\milli\second}.

Following the handshake, the sequences at both nodes enter a photon generation loop.

\paragraph*{Photon generation loop.}
Each iteration of the loop consists of the following operations: 
\begin{enumerate}
	\item Doppler cooling,
		\begin{itemize}
			\item Node A: \SI{63}{\micro\second}
			\item Node B: \SI{60}{\micro\second} + wait time
		\end{itemize}
	\item optical pumping,
		\begin{itemize}
		\item Node A: \SI{280}{\micro\second}
		\item Node B: \SI{60}{\micro\second} + wait time
		\end{itemize}
	\item a bichromatic Raman laser pulse,
		\begin{itemize}
		\item Node A: \SI{50}{\micro\second}
		\item Node B: \SI{50}{\micro\second}
		\end{itemize}
	\item a wait time for a signal that heralds coincident photon detection to be received at both nodes.  
\end{enumerate}
Each iteration lasts \SI{420}{\micro\second}. 
The loop is iterated up to \SI{20}{times}. 
In the absence of coincident photon detection within any of the 20 iterations, the intialization and handshake are repeated. In the case of coincident detection of two photons produced within the same iteration, the loop is terminated, and the sequences proceed to ion-qubit measurement.

\paragraph*{Ion-qubit measurement.}
Measurement of the ion's electronic state at each node proceeds in three steps:
\begin{enumerate}
\item A \SI{729}{\nano\meter} $\pi$-pulse maps the state $\ket{D}\equiv\ket{3^2D_{5/2},m_j=-5/2}$ to $\ket{S}$ at Node A. As a result, information that was encoded in a superposition of $\ket{D}$ and $\ket{D'}\equiv\ket{3^2D_{5/2},m_j=-3/2}$ at each node is now encoded in a superposition of $\ket{S}$ and $\ket{D'}$. At the same time, at Node B, a \SI{729}{\nano\meter} $\pi$-pulse maps the state $\ket{D}\equiv\ket{3^2D_{5/2},m_j=-3/2}$ to $\ket{S}$, so that the encoding is in a superposition of $\ket{S}$ and $\ket{D}$. It is irrelevant whether $\ket{D'}$ or $\ket{D}$ is used for the measurement encoding; the experimenters at the two nodes just happened to make different choices.
	\begin{itemize}
	\item Node A: \SI{5.2}{\micro\second}
	\item Node B: \SI{11.1}{\micro\second}
	\end{itemize} 
\item An optional \SI{729}{\nano\meter} $\pi/2$-pulse is implemented on the $\ket{S}$ to $\ket{D'}$ transition at Node A and on the $\ket{S}$ to $\ket{D}$ transition at Node B~\cite{Haeffner2008}. The pulse is implemented when the ion-qubit is to be measured in the Pauli $\sigma_x$ or $\sigma_y$ basis; we set the optical phase of the pulse to determine in which of the two bases the measurement is made. The pulse is not implemented when the ion-qubit is to be measured in the $\sigma_z$ basis. 
	\begin{itemize}
	\item Node A: \SI{4.3}{\micro\second}
	\item Node B: \SI{7.81}{\micro\second}
\end{itemize} 
\item A projective fluorescence measurement on the \SI{397}{\nano\meter} $4^2S_{1/2} \leftrightarrow 4^2P_{1/2}$ transition determines whether the ion is in $\ket{S}$ or $\ket{D'}$ at Node A, and whether it is in $\ket{S}$ or $\ket{D}$ at Node B. A photomultiplier tube is used to collect fluorescence.
	\begin{itemize}
	\item Node A: \SI{1.5}{\milli\second}
	\item Node B: \SI{1.5}{\milli\second}
\end{itemize} 
\end{enumerate}

\section{Ion-ion state fidelities}
\label{sec:ion-ion_f}

In this section, we explain how uncertainties are calculated for the ion--ion state fidelities presented in the main text.

As described in the main text,  the joint state of two remote ions is characterized via quantum state tomography, yielding the density matrices $\rho^{\pm}(T)$, where $\rho^+$ and $\rho^-$ are reconstructed for the coincidences that should herald the Bell states $|\Psi^+\rangle$ and $|\Psi^-\rangle$, respectively, and $T$ is the maximum time difference between coincident photons for which entanglement is heralded.
The state $\rho^+$ is obtained if coincident detection occurs in the output path of the beamsplitter in which the SNSPDs are placed, while $\rho^-$ is obtained if coincident detection occurs in opposite beamsplitter outputs, i.e., for the two combinations of a coincidence at one SPCM and one SNSPD.
The fidelity is determined according to the expression $F^{\pm}(T) \equiv \langle \Psi^{\pm}| \rho^{\pm}(T) | \Psi^{\pm} \rangle $.

We use Monte Carlo resampling~\cite{Efron93} to obtain the uncertainties in $F^{\pm}(T)$:
Recall that $\rho^{\pm}(T)$ is determined from a set of measurement outcomes, which we can express as a vector. It is assumed that noise on these measurement outcomes is due to projection noise.
We numerically generate $M = 200$ vectors of ``noisy" observations based on a multinomial distribution around the experimentally recorded values.
For each of these vectors, we reconstruct a density matrix just as for the experimental data, via the maximum likelihood technique.
As a result, for each state $\rho^{\pm}(T)$ reconstructed directly from the raw data, we have $M$ states reconstructed from simulated data.
We calculate the value of some quantity of interest, e.g., the fidelity $F^{\pm}(T)$, not only for $\rho^{\pm}(T)$ but also for the associated $M$ states, yielding a distribution $D$ of values with mean $F_{m}$ and standard deviation $\delta$. The uncertainties are then given by $F^{\pm}(T)^{+(F_m+\delta-F)}_{-(F-Fm+\delta)}$.

If $F^{\pm}(T)$ is optimized over the phase $\phi$ of the Bell state, then this calculation is carried out for each value of $\phi$.

\section{Ion-photon state fidelities}
Here, we provide more details on the calibration measurement of ion--photon entanglement that was carried out at each node immediately prior to ion--ion entanglement.

For the ion--photon state generated at Node~B, photons were analyzed using the PBSM setup, details of which are given in Sec.~\ref{sec:PBSM}. Specifically, photon counts were recorded on the two SNSPDs. For the ion--photon state generated at Node~A, photons were analyzed using a separate setup in the Node~A laboratory.
 
For each ion--photon state, measurements are made in all nine combinations of the Pauli measurement bases for two qubits~\cite{James2001}. The measurement basis of the photon is changed using waveplates in the photon analysis path. 
Tomographic reconstruction via the maximum likelihood technique yields the ion--photon density matrices $\rho^{\rm ion-photon}_k$ for $k \in \{\text{A},\text{B}\}$. The fidelities given in the main text are calculated as $\langle \Psi^{\theta}_k | \rho^{\rm ion-photon}_k | \Psi^{\theta}_k \rangle$, where $\ket{\Psi^{\theta}_k} = 1/\sqrt{2}\left (|\text{DV}\rangle  + e^{\mathrm{i} \theta} |\text{D}'\text{H}\rangle\right)$ is the maximally entangled two-qubit state nearest to the state $\rho^{\rm ion-photon}_k$, found by numerical optimization over $\theta$.

The method used to determine uncertainties in these fidelities is described in Sec.~\ref{sec:ion-ion_f}, where we replace the vector of ion--ion measurement outcomes by a vector of ion--photon measurement outcomes. \\

\section{Empirical model for the ion--ion density matrix}
\label{Empirical_model}
The target states for ion--ion entanglement are the two Bell states in Eq.~1 of the main text:
\begin{equation} 
	|\Psi^{\pm} \rangle = 1/\sqrt{2} \left( \ket{\rm D_A D'_B}  \pm e^{\ii \phi} \ket{\rm D'_A D_B} \right).
	\label{eq:eq1s}
\end{equation}
The corresponding density matrices are $\rho^\pm = \ket{\Psi^{\pm}}\bra{\Psi^{\pm}}$. Here we describe an empirical model for the density matrix $\rho$ heralded by two-photon detection in our experiments. For this model, we adapt $\rho^\pm$ to account for three sources of infidelity: detector background counts, photon distinguishability
due to spontaneous emission, and imperfect ion–photon entanglement.

We first account for detector background counts. 
We define $p_{mn}$ as the probability to detect the ion at Node~A in state $m$ and the ion at Node~B in state $n$ in a single experimental trial, where $m, n \in \{\text{D},\text{D}'\}$.
In the absence of detector background counts and all other imperfections, $p_\text{DD} = p_{\text{D}'\text{D}'} = 0$. We write 
\begin{align}
p_\text{DD} &= 
	p_{\rm ph-bg}/4 +p_{\rm bg-bg}/4, \nonumber	\\
p_{\text{D}'\text{D}'} &= p_\text{DD},  \nonumber \\
p_{\rm DD'} &= p_{\rm ph-ph}/2 + p_{\rm ph-bg}/4 +p_{\rm bg-bg}/4,\nonumber\\
p_{\rm D'D} &= p_{\rm DD'},
\label{eq:bkgd_model}
\end{align}
where $p_{\rm ph-bg}$, $p_{\rm bg-bg}$, and $p_{\rm ph-ph}$ are the probabilities to detect a coincidence in a single experimental trial between a photon and a background count, between two background counts, and between two photons. The scaling factors account for the chance to measure a certain ion-ion correlator given a coincidence. An underlying assumption of Eqs.~\eqref{eq:bkgd_model} is that when a coincidence due to one or two background counts occurs, it is equally likely to find the two ions in each of their four possible states. This assumption is valid for the Bell states considered here, and it will still be valid when we introduce a depolarizing channel to model imperfect ion-photon entanglement later in this section.

The ion--ion density matrix that accounts for background counts is given by
\begin{widetext}
\begin{align}
	\rho_{\rm bg}^{\pm} = \frac{1}{\sum_{m,n} p_{mn}} ~~	\begin{blockarray}{ccccc}
		\ket{\rm D'_A D'_B} & \ket{\rm D'_A, D_B} & \ket{\rm D_A, D'_B} & \ket{\rm D_A, D_B}  \\
		\begin{block}{(cccc)c}
			p_{\rm D'D'} & 0 & 0 & 0 &  \:\bra{\rm D'_A, D'_B}\\
			0 & p_{\rm D'D} & \pm e^{-\ii\phi}p_{\rm ph-ph}/2 & 0 &  \:\bra{\rm D'_A, D_B}\\
			0 & \pm e^{\ii\phi}p_{\rm ph-ph}/2 & p_{\rm DD'} & 0 &  \:\bra{\rm D_A, D'_B}\\
			0 & 0 & 0 & p_{\rm DD} &  \:\bra{\rm D_A, D_B}\\
		\end{block}
	\end{blockarray}
\end{align}
\end{widetext}

The matrix $\rho_{\rm bg}$ can also be expressed as 
	\begin{equation}
		\rho_{\rm bg}^{\pm} = \frac{1}{\sum_{m,n} p_{mn}} \left( \frac{p_{\rm ph-ph}}{2} \rho^\pm+\frac{p_{\rm tot-bg}}{4}\mathbbm{1}\right),
	\end{equation}
 where $p_{\rm tot-bg}=p_{\rm ph-bg} +p_{\rm bg-bg} $ and $\mathbbm{1}$ is the two-qubit identity matrix. Here one sees more clearly that the background-count model acts to add white noise to the ion-ion state.

In general, for a given detector combination $\rm det_1$ and $\rm det_2$, one can write the coincidence probabilities as: 
\begin{align} 
	p_{\rm ph-ph} &= p^{\rm A}_{\rm det_1} \times p^{\rm B}_{\rm det_2} + p^{\rm B}_{\rm det_1} \times p^{\rm A}_{\rm det_2}\nonumber\\
	p_{\rm ph-bg} &= (p^{\rm A}_{\rm det_1}+p^{\rm B}_{\rm det_1})\times p_{\rm bg-det_2}\nonumber\\&+
	(p^{\rm A}_{\rm det_2}+p^{\rm B}_{\rm det_2})\times p_{\rm bg-det_1}\nonumber\\
	p_{\rm bg-bg} &= p_{\rm bg-det_1} \times p_{\rm bg-det_2}
	\label{eq:eq4s}
\end{align}
where  $p^{k}_{{\rm det}_r}$ is the probability of detecting a photon at detector $r$ emitted by node $k \in \{\text{A},\text{B}\}$ and $p_{{\rm bg-det}_r}$ is the probability to get a background count within the detection window at detector $r$.
Note that we use four detectors, two of which are SNSPDs and two of which are SPCMs. Background counts and efficiencies have been measured independently for each detector (Sec.~\ref{sec:PBSM}), from which we calculate the probabilities in Eq\.~\eqref{eq:eq4s}.

Second, we account for photon distinguishability 
using a two-qubit dephasing channel. We define a completely dephased density matrix, for which we set the off-diagonal elements of $\rho_{\rm bg}$ to zero:
	\begin{align}
		\rho_{\rm bg,dephase} = \frac{1}{\sum_{m,n} p_{mn}}	\begin{pmatrix}
				p_{\rm D'D'} & 0 & 0 & 0 & \\
				0 & p_{\rm D'D} & 0 & 0 \\
				0 & 0 & p_{\rm DD'} & 0   \\
				0 & 0 & 0 & p_{\rm DD}
		\end{pmatrix}
	\end{align}

The probability for dephasing in the channel
is parameterized by the Hong-Ou-Mandel interference
visibility $V$. The density matrix $\rho_{\rm dist}^{\pm}$ accounts for both background counts and photon distinguishability:
\begin{equation}
	\rho_{\rm dist}^{\pm} = V \times\rho_{\rm bg}^{\pm}
	+ (1 - V) \times \rho_{\rm bg,dephase}.
	\label{eq:rho_dist}
\end{equation}
As discussed in the main text, the value of $V$ is experimentally determined as a function of the coincidence window for photon detection. In the absence of background counts or other imperfections, Eq.~\eqref{eq:rho_dist} predicts an ion--ion state of the form $\rho^{\pm}(1+V)/2 + \rho^{\mp}(1-V)/2$. An equivalent model of the effect of photon distinguishability on entanglement swapping is derived in the Supplemental Material of Ref.~\cite{Craddock2019}; see in particular Eq.~(S29).

Finally, we account for imperfect ion--photon entanglement at Nodes A and B, for which we introduce a two-qubit depolarizing channel.
We define $F'_{\rm ip,k}$ to be the fidelity of ion--photon entanglement with respect to a maximally entangled state at node $k$, where $F'_{\rm ip,k}$ has been corrected for background counts, 
and we define $\rho_{\rm depol}$ to be a completely depolarized density matrix:
\begin{align}
	\rho_{\rm depol} =	\frac{1}{4} \mathbbm{1}.
\end{align}
If we assume that the infidelity $1-F'_{\rm ip,k}$ is due to depolarizing noise, and that the entanglement-swapping process that creates ion--ion entanglement between Nodes A and B is perfect, then the fidelity of ion--ion entanglement with respect to a maximally entangled state is given by \cite{Briegel2000}
\begin{equation}
	F'_{\rm ii} = \frac{1}{4} \left(
1 + 3\left( \frac{4F'_{\rm ip,A}-1}{3} \right) \left( \frac{4F'_{\rm ip,B}-1}{3}
\right)
\right).
\end{equation}
We can then describe the depolarizing channel that generates the state with fidelity $F'_{\rm ii}$ with a parameter $\lambda$ \cite{Horodecki99}: 
\begin{equation}
	\rho^\pm =  \lambda \times \rho_{\rm dist}^{\pm}
	+ (1 -\lambda) \times \rho_{\rm depol},
\end{equation}
where
\begin{align}
	\lambda = \frac{4	F'_{\rm ii}-1}{3}.
\end{align}

We thus arrive at the density matrix $\rho$ from which the fidelities plotted in Fig.~2d of the main text are calculated:
\begin{equation}
	F^\pm_{\rm model} = \bra{\Psi^\pm} \rho^\pm  \ket{\Psi^\pm}.
\end{equation}
In Fig.~2d, the fidelities are plotted as a function of coincidence window. To calculate $\rho$ for a given coincidence window $T$, we take into account the visibility $V(T)$ and the background counts that (on average) occur within the detection window. The depolarizing correction is treated as independent of $T$. To calculate the dashed lines in Fig.~2d, we omit the second step in this model---the dephasing channel parameterized by the visibility--- and determine $\rho$ only taking into account detector background counts and imperfect ion--photon entanglement.

The ion--photon entanglement fidelities at Nodes~A and~B without background-count subtraction are given in the main text. After background-count subtraction, these values are $F'_{\rm ip,A} = (93.8+0.4-0.5)\%$ and $F'_{\rm ip,B} = (95.6+0.7-0.8)\%$.

\section{Master-equation model for two-photon interference visibility}

\subsection{The master equation}\label{sec:Model-master}
We present in this section the master-equation model of the ion--cavity system.  We start with the Hamiltonian, then review the noise terms, and conclude the section with the master equation that is relevant for the description of the experiment. 

Ultimately the model is used to predict the visibility of the interference obtained by combining on a beam-splitter two photons emitted from the two nodes of the ion-trap quantum network (Fig.~3b of the main text). As a first step, we calculate the joint ion--photon states produced at each node. Then the ions are traced out and the interference visibility is computed from the marginal states of the two photons.

\subsubsection{Hamiltonian of the bichromatic cavity-mediated Raman transition}
\label{sec:H-four}
\begin{figure}[h]
 \centering
 \includegraphics[width=0.3\textwidth]{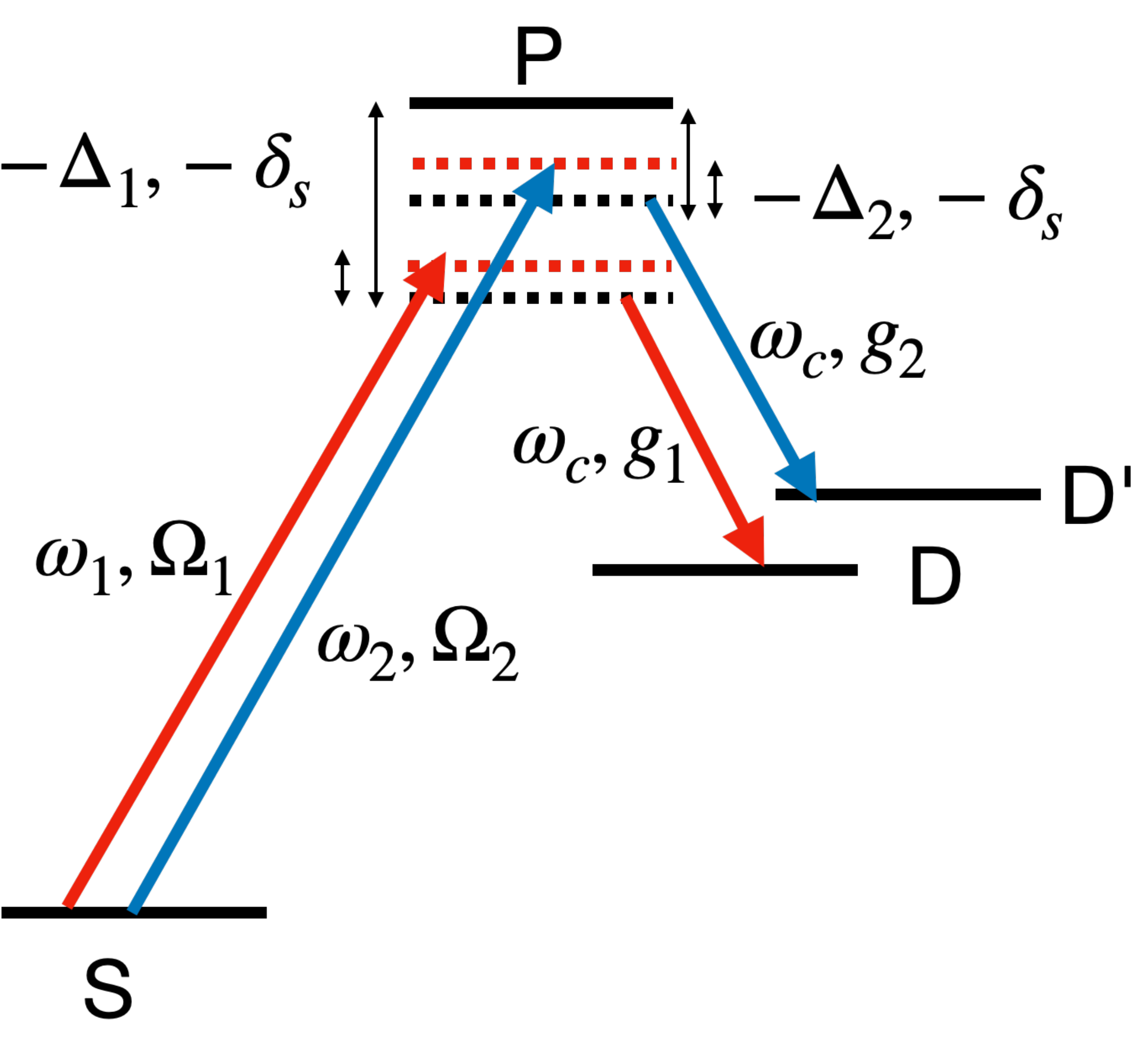}
 \caption{Representation of the energy levels $\ket{S}, \ket{P}, \ket{D},$ and $\ket{D'}$ relevant for the experiment. The frequencies of the bichromatic laser field are denoted $\omega_{1}$ and $\omega_{2}$, with corresponding Rabi frequencies $\Omega_{1}$ and $\Omega_{2}$ and detunings $\Delta_{1}$ and $\Delta_{2}$ from $\ket{P}$. The cavity frequency is $\omega_{\rm c}$, and $g_1$ and $g_2$ are the cavity coupling constants. The Stark shift due to the bichromatic field is $\delta_{\rm s}$. (Note that in Fig.~1 of the main text, $\delta_s$ is set to zero for simplicity.)
  \label{fig:FourLevel}}
 \end{figure}

We start by presenting our model for a single $^{40}$Ca$^+$ ion trapped inside a cavity and driven by laser light. We restrict the atom model to a simple four-level system
that includes the sublevels of direct importance for the experiment: $\ket{S}, \ket{P}, \ket{D},$ and $\ket{D'}$ 
(see Fig.~\ref{fig:FourLevel}). The ion is initially prepared in $\ket{S}$. The $\ket{S}-\ket{P}$ transition is driven off-resonantly with a bichromatic laser field with frequencies $\omega_1$ and $\omega_2$ and Rabi frequencies $\Omega_1$ and $\Omega_2$. The bichromatic field is detuned from the $\ket{S}-\ket{P}$ transition frequency $\omega_{PS}$ by $\Delta_{1}=\omega_{1}-\omega_{PS}$ and $\Delta_{2}=\omega_{2}-\omega_{PS}$. 
In addition, an exchange interaction between the ion and the cavity couples the  $\ket{P}-\ket{D}$ transition to the emission and absorption of a photon with vertical polarization into the cavity and the $\ket{P}-\ket{D'}$ transition to the emission and absorption of a photon with horizontal polarization. The cavity has frequency $\omega_{\rm c}$. The vertically and horizontally polarized cavity modes are described with bosonic operators $\hat a^\dag_{\rm v}$ or $\hat a^\dag_{\rm h}$, and the corresponding coupling constants are denoted $g_1$ and $g_2$.  

\begin{widetext}
The Hamiltonian $H$ of the ion-cavity system is given by
\be
\begin{split}
H/\hbar&=\omega_{\rm c}(\hat  a^\dag_{\rm h} \hat  a_{\rm h}+ \hat a^\dag_{\rm v} \hat  a_{\rm v})+\omega_{PS}\ketbra{P}{P}+\omega_{DS}\ketbra{D}{D}+\omega_{D'S}\ketbra{D'}{D'} \\
&+\frac{1}{2}\Big(\Omega_1e^{\ii\omega_1 t}+\Omega_1e^{-\ii\omega_1 t}\Big)\Big(\ketbra{S}{P}+\ketbra{P}{S}\Big)+\frac{1}{2}\Big(\Omega_2e^{\ii\omega_2 t}+\Omega_2e^{-\ii\omega_2 t}\Big)\Big(\ketbra{S}{P}+\ketbra{P}{S}\Big) \\
&+g
_1\Big(\ketbra{D}{P}+\ketbra{P}{D}\Big)\left(\hat  a^\dag_{\rm v}+\hat  a_{\rm v}\right)+g_2\Big(\ketbra{D'}{P}+\ketbra{P}{D'}\Big)\left(\hat  a^\dag_{\rm h}+\hat  a_{\rm h}\right).
\end{split}
\ee
Note that the energies of the ion levels are defined with respect to $\ket{S}$. An effective Hamiltonian with a simpler form can be obtained by noting that the cavity is initially empty and consequently, the atom-cavity system remains in the four level manifold $\{|S,0\rangle, |P,0\rangle, |D,1_{\rm v}\rangle, |D',1_{\rm h}\rangle\}$, where 0 and 1 are cavity photon numbers and subscripts indicate polarization. The corresponding Hilbert space is labelled $\cH^C$. Below, we shorten the notation to $\ket{D,1_{\rm v}}=\ket{D,1}$ and $\ket{D',1_{\rm h}}=\ket{D',1}$ as there is no ambiguity with the polarization of the cavity photon. Under the rotating wave approximation, the effective Hamiltonian $H_t^C$ is given by
\be
\begin{split}
H_t^C/\hbar&= -\Delta_1 \ketbra{P,0}{P,0}+\big(\Delta_{\rm c_1}-\Delta_1 \big)\ketbra{D,1}{D,1}+\big(\Delta_{\rm c_2}-\Delta_1\big)\ketbra{D',1}{D',1} \\
&+\frac{1}{2}\Big(\Omega_1+\Omega_2e^{\ii(\omega_2-\omega_1)t}\Big)\ketbra{S,0}{P,0}+\frac{1}{2}\Big(\Omega_1+\Omega_2e^{-\ii(\omega_2-\omega_1)t}\Big)\ketbra{P,0}{S,0} \\
&+g_1\Big(\ketbra{D,1}{P,0}+\ketbra{P,0}{D,1}\Big)+g_2\Big(\ketbra{D',1}{P,0}+\ketbra{P,0}{D',1}\Big).
\end{split}
\ee
In the rotating frame $\ket{P}_{L.F.}\rightarrow e^{\ii\omega_1 t}\ket{P}_{R.F.}$, $\ket{1}_{L.F.}\rightarrow e^{\ii\omega_c t}\ket{1}_{R.F.}$, $\ket{D}_{L.F.}\rightarrow e^{\ii(\omega_1-\omega_c)t}\ket{D}_{R.F.}$, and $\ket{D'}_{L.F.}\rightarrow e^{\ii(\omega_1-\omega_c)t}\ket{D'}_{R.F.}$, where $L.F.$ and $R.F.$ stand for lab frame and rotating frame. Here, we have introduced the cavity detunings $\Delta_{\rm c_{1}} = \omega_{\rm c} -\omega_{PD}$ and $\Delta_{\rm c_{2}} = \omega_{\rm c} -\omega_{PD'}$, with $\omega_{PD}= \omega_{PS}-\omega_{DS}$ and $\omega_{PD'}= \omega_{PS}-\omega_{D'S}$. 
In the subspace $\cH_E$ spanned by $\{\ket{D,0},\ket{D',0}\}$, the Hamiltonian is simply
\be\label{eq:HE}
H_E/\hbar = \big(\Delta_{\rm c_1}-\Delta_1 \big)\ketbra{D,0}{D,0}+\big(\Delta_{\rm c_2}-\Delta_1\big)\ketbra{D',0}{D',0}.
\ee

In the experiment, the detunings are calibrated with respect to the observed resonance frequency. It is thus natural to define the detunings $\Delta_{1}' = \Delta_{1}-|\delta_S|$ and $\Delta_{2}' = \Delta_{2}-|\delta_S|$ that incorporate the AC Stark shift $\delta_s=\Omega_1^2/(4\Delta_1)+\Omega_2^2/(4\Delta_2$) calculated for the $\ket{S}-\ket{P}$ transition. In terms of the new detunings, the  Hamiltonian is recast to
\be
\begin{split}
H_t^C/\hbar&=-(\Delta_1'+|\delta_{s}|)\ketbra{P,0}{P,0}+\big(\Delta_{\rm c_1}-\Delta_1'-|\delta_{s}|\big)\ketbra{D,1}{D,1}+\big(\Delta_{\rm c_2}-\Delta_1'-|\delta_{s}|\big)\ketbra{D',1}{D',1} \\
&+\frac{1}{2}\Big(\Omega_1+\Omega_2e^{\ii(\omega_2-\omega_1)t}\Big)\ketbra{S,0}{P,0}+\frac{1}{2}\Big(\Omega_1+\Omega_2e^{-\ii(\omega_2-\omega_1)t}\Big)\ketbra{P,0}{S,0} \\
&+g_1\Big(\ketbra{D,1}{P,0}+\ketbra{P,0}{D,1}\Big)+g
_2\Big(\ketbra{D',1}{P,0}+\ketbra{P,0}{D',1}\Big),\\
H_E/\hbar &=  \big(\Delta_{c_1}-\Delta_1'-|\delta_{s}| \big)\ketbra{D,0}{D,0}+\big(\Delta_{c_2}-\Delta_1'-|\delta_{s}|\big)\ketbra{D',0}{D',0}.
\label{eq:HE2}
\end{split}
\ee
\bigskip
\noindent
The total Hamiltonian is denoted $H_t=H_t^C+H_E.$ 
\end{widetext}

\subsubsection{Noise terms}
 In addition to the Hamiltonian evolution, there are noise terms that affect the dynamics of the system. We review them below.
 
\paragraph* {Spontaneous decay of the ion.}To account for spontaneous decay of the $P$ level to $S$, $D$ or $D'$ (outside of the cavity mode), we introduce the noise operators 
\be
\begin{split}
L_{sp}&= \sqrt{2\gamma_{sp}} \ketbra{S,0}{P,0}, \\ 
L_{dp}&= \sqrt{2 \gamma_{dp}}\ketbra{D,0}{P,0}, \\ 
L_{d'p}&= \sqrt{2 \gamma_{d'p}}\ketbra{D',0}{P,0}, \\
\end{split}
\ee
where $\gamma_{sp}$, $\gamma_{dp}$, and $\gamma_{d'p}$ are atomic polarization decay rates.
These operators pick a phase in the rotating frame. However, these phases do not influence the master equation (see Eq.~\eqref{eq:ME-four-F}) and can thus be ignored.

\paragraph*{Laser noise.}A finite coherence time of the Raman drive laser can be modelled by a process in which each of the Rabi frequencies $\Omega_1$ and $\Omega_2$ (which originate from the same laser field) has a small chance to acquire a random phase $e^{\ii \varphi_t}$ at each moment of time. Since the level $\ket{S,0}$ only couples to other levels by absorbing a laser photon, the laser phase noise can be accounted for in the master equation by introducing a dephasing channel that reduces the coherences $\ketbra{S,0}{P,0}$, $\ketbra{S,0}{D,1}$, and $\ketbra{S,0}{D',1}$. This is done by introducing the noise operator
\be
L_{ss}= \sqrt{2\gamma_{ss}} \ketbra{S,0}{S,0}.
\ee
\paragraph*{Cavity jitter.}
The cavity jitter stems from slow drifts of the cavity frequency away from the reference frequency between recalibration steps, which we attribute to imperfect active stabilization of the cavity length. The resonator is a massive system, so that the cavity length drifts on timescales much slower than the duration of the Raman pulse. Therefore, we assume the cavity frequency $\omega_{\rm c}$ to be fixed during a single iteration of the experiment (i.e., an attempt to generate a single photon). On the other hand, $\omega_{\rm c}$ can change from one iteration to the next. We thus assume that for each iteration, the cavity frequency is a Gaussian random variable with standard deviation $\gamma_{clj}$, which is well justified because the data analysis of the run sequence is unordered. That is, at each iteration, $\hat{\omega}_{\rm c}$ is sampled from the Gaussian distribution
\be
\label{eq:Gaussian}
\text{p}(\hat {\omega}_{\rm c})  
= \frac{1}{\sqrt{2\pi }} \exp\left (-{\frac{(\hat{\omega}_{\rm c}-\omega_{\rm c})^2}{2 \gamma_{clj}^2}} \right).
\ee
Concretely, this means that we solve the dynamics of the two ion--cavity systems for fixed values of $\hat{\omega}_{\rm c}$ that are sampled from $\text{p}(\hat {\omega}_{\rm c})$. The final state is a mixture of these solutions. In practice, to compute the model for $\hat \omega_{\rm c}$, we take a discrete ensemble of $2 k_{\rm max} +1$ equally spaced values $\omega_k = w_{\rm c} + \Delta k$ for $|k| \leq k_{\rm max}$, then renormalize the distribution by a constant such that it sums to one: $\sum_k \text{p}(\omega_k) =1$, that is, the contribution of each frequency in the ensemble is weighted by the distribution. For the numerical analysis below, we take $k_{\rm max}=6$ for Node A (yielding $13$ possible values for $\hat \omega_{\rm c}$), and we neglect the effect of the cavity lock jitter for Node B (fixing $\hat \omega_c = \omega_c$) as it is estimated to be an order of magnitude smaller.

\paragraph*{Photon emission}
The possibility for the photon to leave the cavity gives rise to two noise operators 
\be\begin{split}
	\label{eq:L4L5}
L_4 &= \sqrt{2\kappa} \ketbra{D,0}{D,1} \\
L_5 &=  \sqrt{2\kappa} \ketbra{D',0}{D',1}
\end{split}
\ee
with $\kappa$ the cavity field decay rate.
In our rotating frame, the noise operators are time dependent: $L_4 = \sqrt{2\kappa} \ketbra{D,0}{D,1} e^{-\ii \omega_c t}$, $L_5 = \sqrt{2\kappa} \ketbra{D',0}{D',1} e^{-\ii \omega_{\rm c} t}$. For the master equation, however (see Eq.~\eqref{eq:ME-four-F}), the phase of the noise operators plays no role. 
Note that the noise channels $L_4$ and $L_5$ encompass all cavity decay processes, including transmission, scattering, and absorption at both mirrors. Only a fraction of these photons are transmitted through the output mirror and sent to the PBSM.

\subsubsection{The master equation for the full dynamics}
\label{sec:four-photon}
To capture not only the unitary dynamics of the ion-cavity system but also decoherence and photon emission from the cavity, we use the master equation 
\be\label{eq:ME-four-F}
\dot \varrho_t = -\ii\, [H_t,\varrho_t]/\hbar + \sum_{i} \left(L_i \varrho_t L_i^\dag - \frac{1}{2}\{L_i^\dag L_i, \varrho_t\}\right) ,
\ee
where the density matrix $\varrho_t$ is defined on the six-level subspace $\cH$ spanned by $\{|S,0\rangle, |P,0\rangle, |D,1\rangle,  |D',1\rangle,|D,0\rangle,|D',0\rangle\}$. The index $i$ includes all the terms described above, that is, 
$i=sp,ss,dp,d'p,4,5$. The probability density (rate) for a noise event $L_i$ to occur at time $t$ is denoted by $P_i(t) = \tr L_i \varrho_t L_i^\dag$. The event leaves the system in the state
\be
\varrho_{t|i} = \frac{L_i \varrho_t L_i^\dag}{\tr L_i \varrho_t L_i^\dag}.
\ee

\subsection{Photon envelope and scattering rates}
In this section, we solve the dynamics of the master-equation model developed in Sec.~\ref{sec:Model-master} for the ion--cavity system. As we will see, it is enough to model the system's state inside the four-dimensional subspace $\cH^C$ for this purpose. Below, the density matrix is thus restricted to this subspace.

Knowledge of the ion--cavity state is sufficient to predict the scattering rates and the temporal envelopes of photons leaving the cavity. Through a comparison between the prediction of our theoretical model and the experimental data for the photon temporal envelopes, we are able to fix free parameters in the model, including the cavity loss, cavity jitter and the overall detection efficiency.

\subsubsection{Ion-cavity dynamics}
In the master equation given in Eq.~\eqref{eq:ME-four-F}, different noises play different roles. The terms $L_{sp}$ and $L_{ss}$ leave the system in a state in the $\cH^C$ subspace where it can still emit a photon. However, if the noise events $L_{dp}$, $L_{d' p}, L_4$, or $L_5$ occur, no photon can be emitted afterwards as the system is projected into $\cH_E$. Since we are only interested in the evolution branch that can lead to the emission of a photon, we solve the master equation with the system remaining inside $\cH^C$, that is,
\be\label{eq:ME-four}\begin{split}
\dot \varrho_t =& -\ii \, [H_t^C,\varrho_t]/\hbar + \sum_{i=sp,ss} \left( L_i \varrho_t L_i^\dag - \frac{1}{2} \{ L_i^\dag L_i, \varrho_t \}\right) \\
&-  \sum_{i=dp,d'p,4,5}  \frac{1}{2} \{L_i^\dag L_i, \varrho_t \}.
\end{split}
\ee
Note that the solution of this equation is not trace preserving, as it ignores the branches where $L_{dp}$, $L_{d' p}, L_4$, or $L_5$ happen. In fact, the trace of $\varrho_t$ gives the probability that none of these noises have happened before time $t$.

\subsubsection{Photon envelope}
We are primarily interested in the emission of a photon from each cavity to the PBSM setup when the ion-cavity system is initially in state $\varrho_0= \ketbra{S,0}$. If a photon is generated in the cavity mode, it leaves the cavity with rate $2\kappa$. To compute the probability that a photon of a given polarization (horizontal or vertical) is emitted at time $t$, it is thus sufficient to solve the master equation~\eqref{eq:ME-four} for the initial state $\varrho_0= \ketbra{S,0}$ and compute
\be\label{eq:envelope}\begin{split}
p_{\rm v}(t) &= 2\kappa \bra{D,1} \varrho_t \ket{D,1}\\
p_{\rm h}(t) &= 2\kappa \bra{D',1} \varrho_t \ket{D',1}.
\end{split}
\ee
The envelope of this photon is thus defined by the functions $p_{\rm v}(t)$ and $p_{
\rm h}(t)$. In the presence of cavity jitter, the photon envelopes are the weighted averages over the different cavity frequency values
\be \label{eq:weighted_envelope} \begin{split}
\bar{p}_{\rm v}(t) =\sum_k \text{p}(\omega_k)\,  p_{\rm v}(t|\omega_k),\\
\bar{p}_{\rm h}(t) =\sum_k \text{p}(\omega_k)\,  p_{\rm h}(t|\omega_k), 
\end{split}
\ee
where $p_{\rm v}(t|\omega_k)$ and $p_{\rm h}(t|\omega_k)$ give the probabilities that a photon of a given polarization leaves the cavity at time $t$ for a fixed cavity frequency $\omega_k$.
The photon envelopes of Eq.~\eqref{eq:envelope} and Eq.~\eqref{eq:weighted_envelope} can be compared with the time histograms of click events obtained at the PBSM setup. For these measurements, data are taken when only one node is sending photons, while the other is blocked.

\begin{figure}[t!]
    \centering
    \includegraphics[width=0.98\columnwidth]{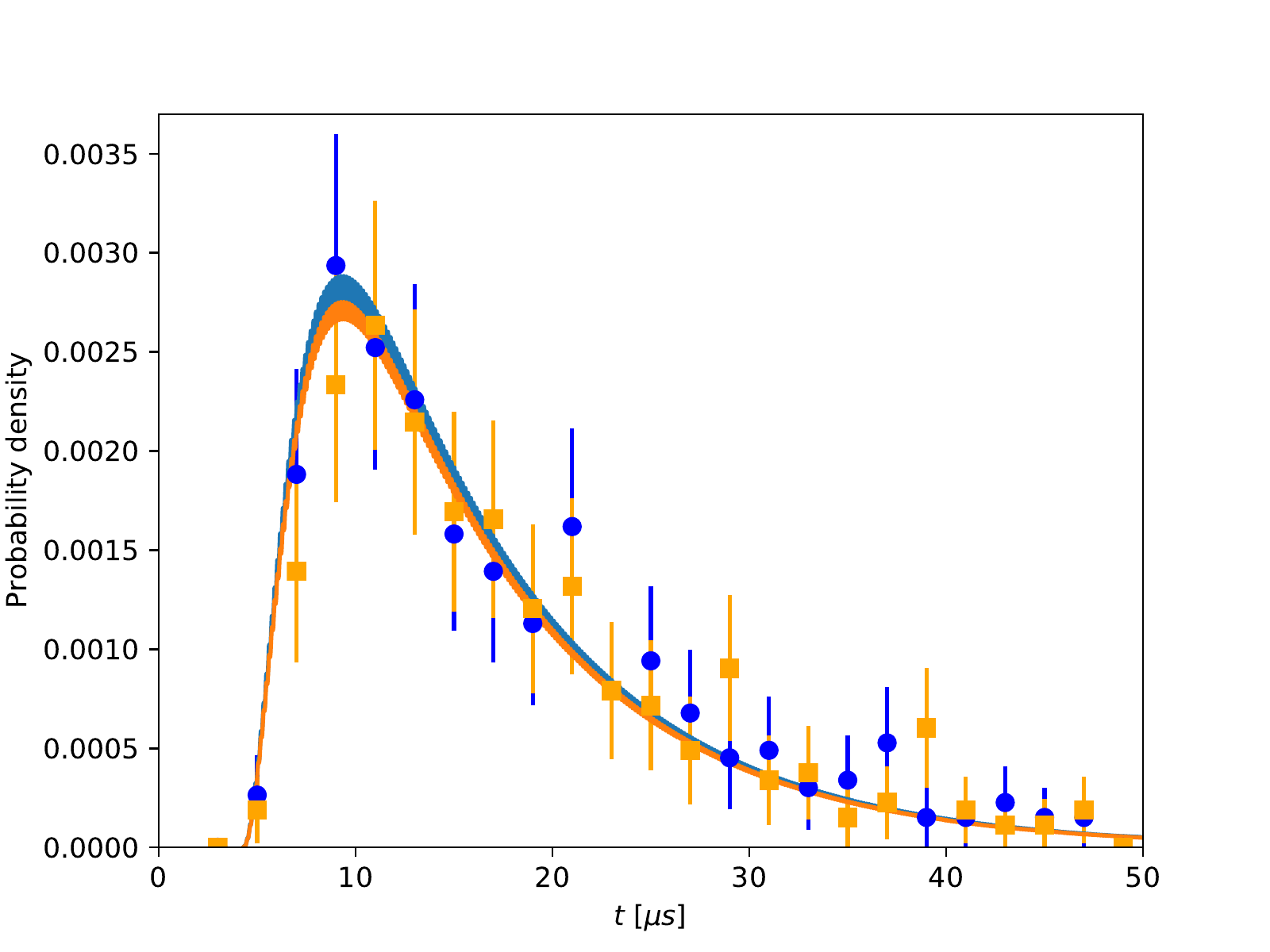}
    \caption{Single-photon temporal wavepacket emitted from Node B and detected on the PBSM setup. 
    Orange squares and blue circles correspond to vertical and horizontal polarizations.
    Squares and circles represent experimental data; error bars are calculated from Poissonian statistics.
    Lines are the envelopes found theoretically, which have been multiplied by $\eta=1/10.5\approx 0.095$.}
    \label{fig:benwavepacket}
\end{figure}

 In Fig.~\ref{fig:benwavepacket}, we compare our model with data obtained from Node B. To obtain agreement between the observed detection rates and the model, we have multiplied the predicted emission rate $\bar{p}_{\rm h(v)}(t)$ by a factor $1/10.5\approx 0.095$, which corresponds to the overall detection efficiency $\eta$, including detector efficiencies, photon loss in the channel, and scattering and absorption losses contained in the noise channels $L_4$ and $L_5$.

\begin{figure}[t!]
    \subfloat{\includegraphics[width=0.98\columnwidth]{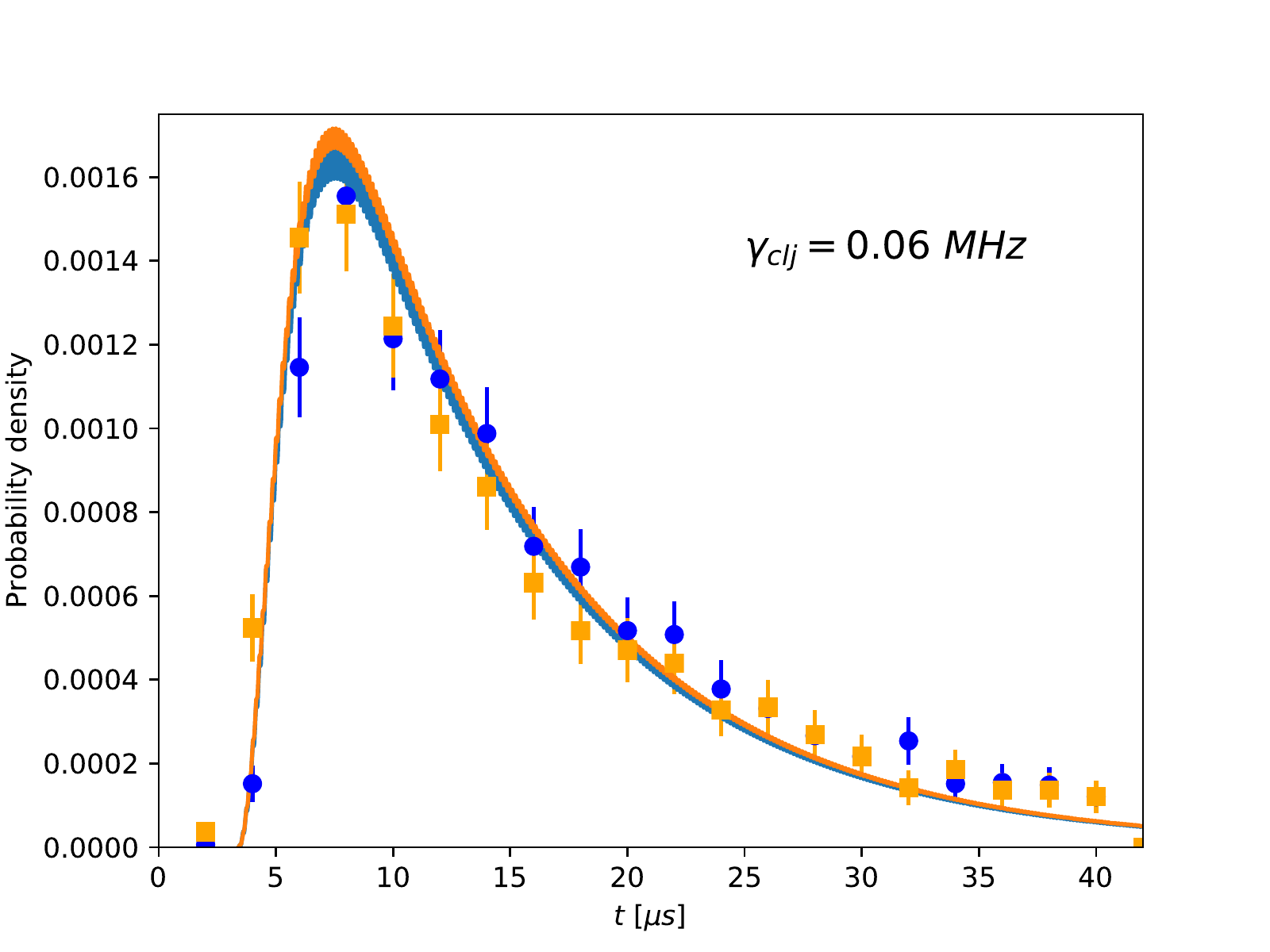}}
	\hspace{0.1cm}
    \subfloat{\includegraphics[width=0.98\columnwidth]{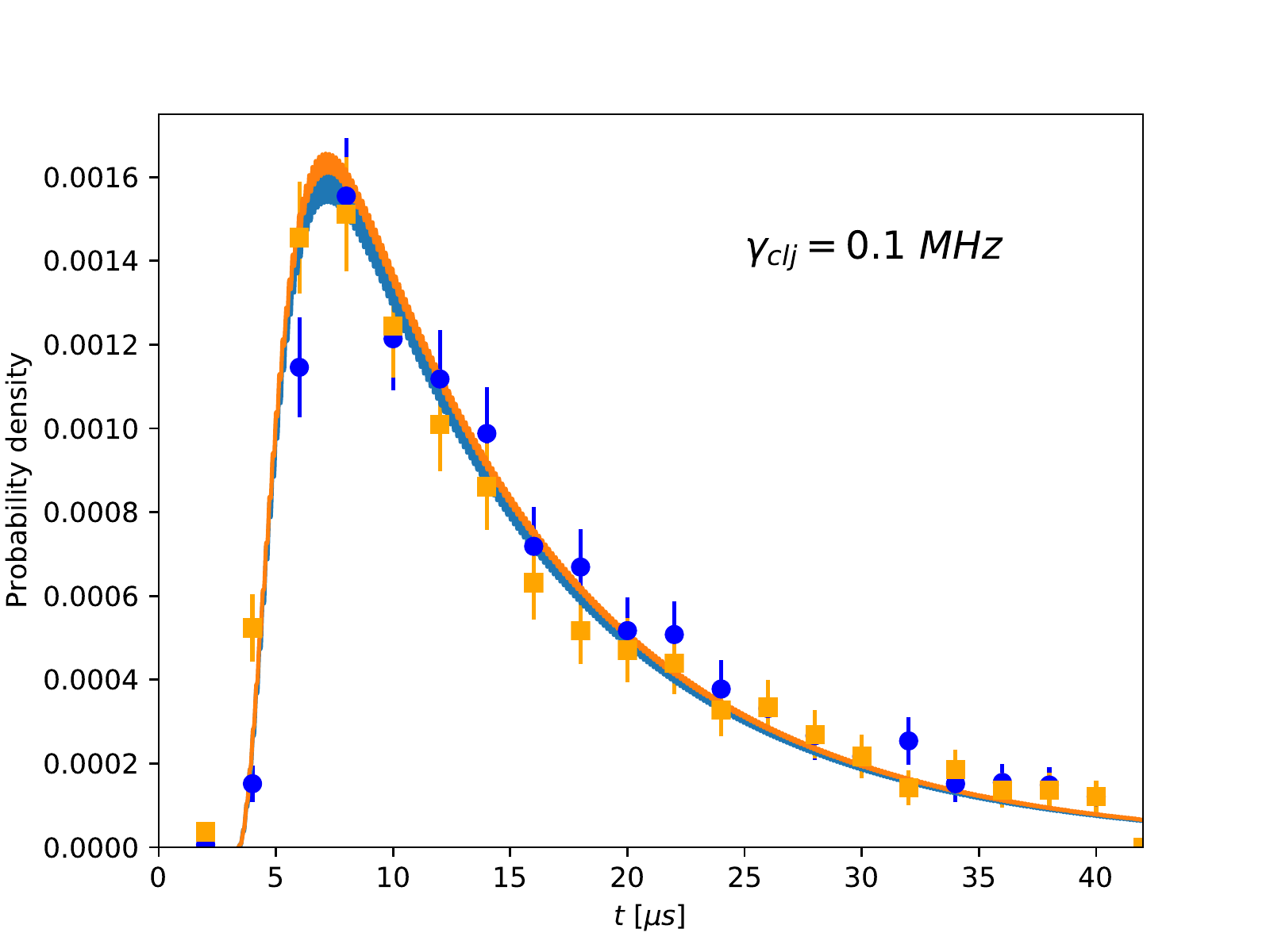}}
    \caption{Single-photon temporal wavepacket emitted from Node A and detected a few meters away. Orange squares and blue circles correspond to vertical and horizontal polarizations. Squares and circles represent experimental data; error bars are calculated from Poissonian statistics. Lines are the envelopes found theoretically, which have been multiplied by $\eta=1/14.47\approx 0.069$ (above) and $\eta =1/12.46\approx0.08$ (below). Cavity jitter has been added with $\gamma_{clj}=0.06$ (above) and $\gamma_{clj}= 0.1$ (below). Both parameter regimes are consistent with the data, that is, are within the uncertainties of experimentally determined values for $\eta$ and $\gamma_{clj}$.}
    \label{fig:tracywavepacket}
\end{figure}

 In Fig.~\ref{fig:tracywavepacket}, we compare our model with data obtained from Node A. Here as well, the predicted emission rates at time $t$ are multiplied by a prefactor that accounts for the detection efficiency. In contrast to the comparison in Fig.~\ref{fig:benwavepacket}, here we include cavity jitter, that is, we use Eq.~\eqref{eq:weighted_envelope} instead of Eq.~\eqref{eq:envelope}. All parameters used for the numerical simulation are reported in Table.~\ref{tab:params}.

\begin{table*}[t]
 \label{table 1}
\centering
\begin{tabular}{|l | c c c c c c c c c c c|} 
\hline 
Node & $\Omega_1$ & $\Omega_2$ & $g$ & $\Delta_1$ & $\Delta_2$ &$\kappa$ & $\gamma_{sp}$ & $\gamma_{dp}+\gamma_{d'p}$ & $\gamma_{ss}$ & $\gamma_{clj}$ &$\eta$\\
 \hline
A & 43.8 & 30.9 & 0.77 & 412.8206 & 419.8574 & 0.0684 & 10.74 & 0.75 & 0.01 & 0.06 -- 0.1 & 0.069-- 0.08\\
 B & 24.76 & 21.05 & 1.2 & 414.0917 & 421.2091 & 0.07 & 10.74 & 0.75 & 0 & 0 & 0.095\\
 \hline
\end{tabular}
\caption{\label{tab:params}  The parameters that are used in the theoretical model to simulate the experimental data. All parameters have units of $\rm MHz$ and must be multiplied by $2\pi$. In order to obtain the coupling strengths $g_1$ and $g_2$ shown in Fig.~\ref{fig:FourLevel}, we multiply $g$ with the relevant atomic transition strength and with the projection of the transition polarization onto the photon polarization \cite{Stute2012a}.} 
\end{table*}

\subsubsection{Scattering rates}
To compute the interference visibility in the next section, we need to predict the scattering rates of the ion-cavity system back to its initial state. Once Eq.~\eqref{eq:ME-four} has been solved and the state $\varrho_t$ has been computed, the rate of scattering back to to $\ket{S,0}$ can be obtained as
\be\label{eq:scatt-Ps}
\text{P}_s(t) = \tr \Big((L_{sp}^\dag L_{sp} +L_{ss}^\dag L_{ss}) \varrho_t \Big).
\ee
Note that whenever such a scattering event occurs, the system is projected onto the state $\ket{S,0}$ at the corresponding time. 

\subsection{The full state of the photon}

The photon envelopes  $p_{\rm v}(t)$ and $p_{\rm h}(t)$ defined in Eq.~\eqref{eq:envelope} give the probabilities for photon emission at different times, but they do not tell us how coherent the emission process is. In particular, they do not tell us about the purity of the state of the emitted photon (for a fixed polarization) and are not sufficient to predict the interference visibility between two photons coming from different nodes. A more detailed analysis is thus required.

Such an analysis is reported below in three steps. First, we compute the ion--cavity state conditional on no noise events occurring during the evolution. Combining this pure state with the scattering rate computed in the previous section, we compute the actual ion--photon state. Finally, tracing out the ion, we obtain the full state of the photon emitted from each cavity and use it to predict the interference visibility.

\subsubsection{No-noise branch}

To compute the final ion--photon state, 
our first step is to extract from the master equation the branch that corresponds to the evolution branch on which no noise events occur. This is given by the equation
\be
\dot{\rho} = -\ii[H_t^C, \rho]/\hbar - \frac{1}{2} \{ \sum_i L_i^\dag L_i,  \rho\},
\ee
where we have simply removed all post-noise terms $L_i \rho_t L_i^\dag$. This equation can be cast in the form
\be\begin{split}
&\dot{\rho_t} = - D_t \rho_t - \rho_t D_t^\dag, \\
&\text{with} \quad
D_t = \ii H_t^C/\hbar +\frac{1}{2} \sum_i L_i^\dag L_i.
\end{split}
\ee
One sees that if the state is initially pure,  $\rho_{t_0}=\ketbra{\Psi_{t_0}}$, it will remain pure in the no-noise-branch evolution, that is, the evolution given by the Schr\"odinger equation 
\be \label{eq: non-hermitian shrodinger}
\ket{\dot{\Psi}_{t}} = -D_t  \ket{\Psi_{t}},
\ee
with a non-Hermitian Hamiltonian $D_t$. The norm of the state decreases in general as $\frac{d}{dt}\| \ket{{\Psi}_{t}}\| = - \bra{{\Psi}_{t}} \sum_i L_i^\dag L_i \ket{{\Psi}_{t}}$, reflecting the fact that the system leaves the no-noise branch whenever a noise event occurs.
The solution of Eq.~\eqref{eq: non-hermitian shrodinger} can be expressed formally by defining the time-ordered propagator
\be\begin{split}
 \ket{\Psi_t} &= V_{t_0}(t-t_0)\ket{\Psi_{t_0}}, \\
 V_{t_0}(\tau) &= \mathcal{T} \left [e^{-\int_{t_0}^{t_0+\tau}D_s \dd s} \right],
\end{split}\ee
where $\mathcal{T}[\bullet]$ is the time-ordering operator.

For our noise model, the initial state for the no-noise evolution is always pure and given by $\ket{\Psi_{t_0}} = \ket{S,0}$ for some time $t_0$ where $t_0$ is determined by a noise event projecting the system onto $\ket{S}$, as discussed below. Let us denote $\ket{\Psi_{t|t_0}}$ the state of the system at time $t$, given that it was prepared in $\ket{S,0}$ at time $t_0\leq t$ and no scattering events occurred in between, that is, given that the system has evolved between $t_0$ and $t$ following the no-noise branch. This state is the solution of $\ket{\dot{\Psi}_{t|t_0}} = -D_t  \ket{\Psi_{t|t_0}}$ and can also be expressed as 
\be \label{eq:psitt0}
\ket{{\Psi}_{t|t_0}}= V_{t_0}(t-t_0)\ket{S,0}.
\ee

It is worth noting that the Hamiltonian has a time dependence, meaning that time-translation symmetry is broken: $V_{t_1}(\tau) \neq V_{t_0}(\tau)$, that is, the evolution for a duration $\tau$ depends on the start time. Nevertheless, in our numerical computations we ignore this asymmetry and use the approximation 
 $\ket{\Psi_{t|t_0}} \approx \ket{\Psi_{(t-t_0)|0}}$. This approximation results in a substantial computational speedup. We have established the validity of this approximation by comparing its results with the results of a time-dependent model for several values of $t_0$.

\subsubsection{Ion--cavity state revisited}
\label{sec:ion_cavity_revisited}

At this point, we know how to compute the scattering rate $\text{P}_s(t)$ and the state $\ket{\Psi_{t|t_0}}$. It is then convenient to express the total state of the system in the form
\be \label{eq: state decomposition}
\begin{split}
\varrho_t &=  \ketbra{\Psi_{t|0}}  + \int_{0}^t \dd s\,  \text{P}_s(s)\,  \ketbra{\Psi_{t|s}}\\
& \approx  \ketbra{\Psi_{t|0}}  + \int_{0}^t \dd s\,  \text{P}_s(s)\,  \ketbra{\Psi_{t-s|0}},
\end{split}\ee
where in the second step, we use the approximation $\ket{\Psi_{t|t_0}} \approx \ket{\Psi_{(t-t_0)|0}}$ discussed above. This expression captures the fact that given a state at a certain time, the system will either evolve without noise until $t$ (no-noise branch), trigger a noise event $L_{ss}$ or $L_{ps}$ at a later time $t'$ ($s\leq t'\leq t$) that keeps it within the four-dimensional manifold $\cH^C$, or trigger one of the other four noise events that causes it to leave $\cH^C$ (and never emit a photon that is sent to the PBSM setup). Note that the probability that at time $t$, the most recent scattering event happened at time $s\leq t$ is $\dd s\,  \text{P}_s(s)\|\ket{\Psi_{t|s}}\|$, which explains the term in the integral of Eq.~\eqref{eq: state decomposition}.

\subsubsection{Ion--photon state}
We now show that the decomposition of the state $\varrho_t$ in the form proposed in Eq.~\eqref{eq: state decomposition} results in a natural description of the entangled state of the ion and the cavity photon. First, note that the states entering in the decomposition ($\Psi_{t|s}$) are pure, i.e., Eq.~\eqref{eq: state decomposition}  gives an explicit decomposition of $\varrho_t$ into pure states. For a pure ion--cavity state $\ket{\Psi_{t}}$, the probability amplitude that a photon leaves the cavity after a time duration $\dd t$ (corresponding to the  $L_4$ and $L_5$ decay channels when the photon is traced out) is obtained from
\be\begin{split}
&\dd t \, E_t\ket{\Psi_t} \equiv \\
&\dd t \sqrt{2 \kappa}\left(\ketbra{D,0}{D,1} a_{\rm v}^\dag(t) +\ketbra{D',0}{D',1} a_{\rm_h}^\dag(t)\right) \ket{\Psi_{t}} ,
\end{split}
\ee
where the ion--cavity state is projected into the $\cH_E$ subspace. Here we have introduced the creation and annihilation operators for the continuous temporal (and polarization) modes outside the cavity directed to the PBSM setup, which satisfy $[a_{\rm v}(t),a_{\rm v}^\dag(t')] =[a_{\rm h}(t),a_{\rm h}^\dag(t')]=\delta(t-t')$. Thus, for the ion--cavity system evolving in the no-noise branch,  with the system in state $\ket{S,0}$ at time $s$ and in $\ket{\Psi_{t|s}}$ at time $t$, we can associate a probability amplitude that a photon is emitted from the cavity towards the PBSM setup in an infinitesimal time window $[t',t'+dt']$ with $s\leq t'$  and $t'+\dd t'\leq t$. These events are coherent and described by the states $E_t'\,  \dd t' \ket{\Psi_{t'|s}}\ket{0}_{t'}$, where $\ket{0}_{t'}$ is the vacuum state of all the temporal modes in the interval  $[t',t'+dt']$.
It follows that the no-noise evolution branch  corresponds to a branch where a single photon has been coherently emitted,  which is described by the state
\be\label{eq: emision state}\begin{split}
  & \left( \int_s^t \dd t' \, e^{-\ii(t-t') H_E} E_{t'} \ket{\Psi_{t'|s}}\right) \ket{\bm 0} = \sqrt{2\kappa} \int_s^t \dd t' \times \\
&\Big( 
 \ket{D,0} e^{-\ii (t-t')(\Delta_{{\rm c}_1}-\Delta_1'-|\delta_{s}|)} \braket{D,1}{\Psi_{t'|s}} a_{\rm v}^\dag(t') \\
 &+ \ket{D',0} 
e^{-\ii (t-t')(\Delta_{{\rm c}_2}-\Delta_1'-|\delta_{s}|)}\braket{D',1}{\Psi_{t'|s}} a_{\rm h}^\dag(t') \Big)\ket{\bm 0}.
\end{split}
\ee
Here, $\ket{\bm 0}$ denotes all the temporal modes of the photons traveling to the PBSM setup. 
In Eq.~\eqref{eq: emision state}, the term $ e^{-\ii(t-t') H_E}$ describes the evolution of the ion--cavity system following the emission of a photon at time $t'$. Recall from Eq.~\eqref{eq:HE2} that  the states $\ket{D,0}$ and $\ket{D',0}$ acquire phases  $ \ket{D,0} \mapsto e^{-\ii (t-t')(\Delta_{{\rm c}_1}-\Delta_1'-|\delta_{s}|}\ket{D,0}$ and $\ket{D',0} \mapsto e^{-\ii (t-t')(\Delta_{{\rm c}_2}-\Delta_1'-|\delta_{s}|)}\ket{D',0}$ between the times $t'$ and $t$, as given by the energies of the Hamiltonian $H_E$. To shorten the notation, it is convenient to introduce the complex amplitudes
\be\begin{split}
\alpha(t'|s) &=  e^{\ii t' (\Delta_{\rm c_1}-\Delta_1'-|\delta_{s}|)} \braket{D,1}{\Psi_{t'|s}}, \\
\beta(t'|s) &=  e^{\ii t'(\Delta_{\rm c_2}-\Delta_1'-|\delta_{s}|)} \braket{D',1}{\Psi_{t'|s}}.
\end{split}\ee
Then in the photon-emitted branch of the evolution, with the ion--cavity system prepared in $\ket{S,0}$ at time $s$, the ion--photon state at time $t$ is given by
\be
\label{eq: phi t|s}
\begin{split}
\ket{\Phi_{t|s}} = &\sqrt{2 \kappa}\Big(\ket{D,0} \int_{s}^t \dd t' \alpha(t'|s) a_{\rm v}^\dag(t') \\
&+ e^{\ii t(\Delta_{\rm c_1}-\Delta_{\rm c_2})} \ket{D',0} \int_{s}^t \dd t' \beta(t'|s) a_{\rm h}^\dag(t') \Big) \ket{\bm 0}.
\end{split}\ee
This state can be rewritten as
\be
\label{eq:phits}
\ket{\Phi_{t|s}} = \ket{D,0}\ket{V_{t|s}} + e^{\ii t(\Delta_{\rm c_1}-\Delta_{\rm c_2})} \ket{D',0}\ket{H_{t|s}}
\ee
with the unnormalized single-photon states
\be\begin{split}
\label{eq:htsvts}
&\ket{V_{t|s}} = \sqrt{2 \kappa} \int_{s}^t \dd t' \alpha(t'|s) a_{\rm v}^\dag(t') \ket{\bm 0},   \\
&\ket{H_{t|s}} = \sqrt{2 \kappa} \int_{s}^t \dd t' \beta(t'|s) a_{\rm h}^\dag(t') \ket{\bm 0}.
\end{split}
\ee
From this point on, we will write $\ket{D}$ and $\ket{D'}$ instead of $\ket{D,0}$ and $\ket{D',0}$ since there is no ambiguity concerning the absence of cavity photons.

From the decomposition in pure states of the ion--cavity state given in Eq.~\eqref{eq: state decomposition}, we can now deduce the full (unnormalized) ion--photon state associated with the evolution branch in which a single cavity photon has been emitted towards the PBSM setup:
\be
\label{eq:ion-photon state}
\rho^E_{t} =  \ketbra{\Phi_{t|0}}  + \int_{0}^t \dd s\,  \text{P}_s(s)\,  \ketbra{\Phi_{t|s}},
\ee
with the pure states $\ket{\Phi_{t|s}}$ given in Eqs.~\eqref{eq: phi t|s} and \eqref{eq:phits}.
 
\subsubsection{The marginal state of the photon}
From the ion--photon state $\rho^{E}_t$ (Eq.~\eqref{eq:ion-photon state}) (with an empty cavity), it is straightforward to compute the marginal state $\sigma_t$ of the emitted photon by tracing out the ion--cavity system. We obtain
\be\label{eq: photon states}
\sigma_t = \tr_{\rm ion-cavity} \rho^E_t = \mathbf{V}_t + \mathbf{H}_t,
\ee
with 
\be\label{eq: photon states 2}\begin{split}
	\mathbf{V}_t &=  \ketbra{V_{t|0}}  + \int_{0}^t \dd s\,  \text{P}_s(s)\,  \ketbra{V_{t|s}}, \\
  \mathbf{H}_t &=  \ketbra{H_{t|0}}  + \int_{0}^t \dd s\,  \text{P}_s(s)\,  \ketbra{H_{t|s}}.
\end{split}
\ee
Here the density matrices $\mathbf{V}_t$ and $\mathbf{H}_t$ are not normalized. Their traces corresponds to the probabilities that a vertically or horizontally polarized photon has been emitted outside of the cavity in the mode of interest before time~$t$.

Note that the components of the states in Eq.~\eqref{eq: photon states 2} can be conveniently written as  
\be\begin{split}
&\mathbf{V}_t = \int_{0}^t \widetilde{\text{P}}_s (s) \ketbra{V_{t|s}} \\
 &\text{with~} \widetilde{\text{P}}_s (s) =\text{P}_s(s) + \delta(s)
\end{split}\ee
such that $\int_{0}^t \dd s\, \delta(s) \ketbra{V_{t|s}} = \ketbra{V_{t|0}}$, with an equivalent expression for $\mathbf{H}_t$. It is then straightforward to include the effects of cavity jitter, as we now show.

\subsubsection{Effects of cavity jitter}\label{sec:Eff-jitt}
We first remark that the above derivation of the ion--photon state assumes that the cavity frequency $\omega_{\rm c}$ is constant, which is not the case in  the presence of cavity jitter, where $\hat \omega_c$ is a random variable distributed according to $\text{p}(\delta w)$, as discussed earlier in the context of Eq.~\eqref{eq:Gaussian}. Nevertheless, the effects of cavity jitter on the final state can be straightforwardly included, as we now discuss.

In our model, we take a discrete set of possible values: $\hat{\omega}_c \in \{\omega_k\}_{k=1}^{13}$. The final ion-cavity state is then a mixture 
\be
\bar{\rho}_t^E = \sum_{k}  \text{p}(\omega_k) \rho^{E,(\delta \omega_k)}_t \qquad \text{for} \qquad \delta \omega_k = \omega_k -\omega_c,
\ee
where each state $\rho^{E,(\delta w)}_t$ takes the form 
\be\begin{split}
\rho_{t}^{E,(\delta w)} &=  \ketbra{\Phi^{(\delta w)}_{t|0}}  \\
&+ \int_{0}^t \dd s\,  \text{P}^{(\delta w)}_s(s)\,  \ketbra{\Phi^{(\delta w)}_{t|s}}, \\
\alpha^{(\delta w)}(t'|s) &=  e^{\ii t' (\widehat{\Delta}_{\rm c_1}-\Delta_1'-|\delta_{s}|)} \braket{D,1}{\Psi^{(\delta w)}_{t'|s}} 
\\ &= e^{\ii t' (\Delta_{\rm c_1} +\delta w -\Delta_1'-|\delta_{s}|)} \braket{D,1}{\Psi^{(\delta w)}_{t'|s}}\\
\beta^{(\delta w)}(t'|s) &=  e^{\ii t' (\widehat{\Delta}_{\rm c_2}-\Delta_1'-|\delta_{s}|)} \braket{D',1}{\Psi^{(\delta w)}_{t'|s}}\\
& = e^{\ii t' (\Delta_{\rm c_2} +\delta w -\Delta_1'-|\delta_{s}|)} \braket{D',1}{\Psi^{(\delta w)}_{t'|s}};
\end{split}
\ee
see Eqs.~\eqref{eq:phits} and \eqref{eq:htsvts}. Here $\text{P}^{(\delta w)}_s(s)$ and $\ket{\Psi^{(\delta w)}_{t|s}}$ are obtained similarly to $\text{P}_s(s)$ in Eq.~\eqref{eq:scatt-Ps} and $\ket{\Psi_{t|s}}$ in Eq.~\eqref{eq:psitt0} for a shifted cavity frequency $\omega_c +\delta w$. 

The final state of the emitted photon also becomes a statistical mixture
over the possible values of the cavity frequency $\omega_c+\delta\omega_k$:
\be
\bar{\mathbf{V}}_t = \sum_{k}  \text{p}(\omega_k) \mathbf{V}_t^{(\delta \omega_k)},
\qquad
\bar{\mathbf{H}}_t = \sum_{k}  \text{p}(\omega_k) \mathbf{H}_t^{(\delta \omega_k)},
\ee
with 
\be
\begin{split}
	\mathbf{V}_t^{(\delta \omega)} &=  \int_{0}^t \widetilde{\text{P}}^{(\delta \omega)}_s (s) \ketbra{V_{t|s}^{(\delta \omega)}}, \\
	\mathbf{H}_t^{(\delta \omega)} &=  \int_{0}^t \widetilde{\text{P}}^{(\delta \omega)}_s (s) \ketbra{H_{t|s}^{(\delta \omega)}}, \\
	\ket{V_{t|s}^{(\delta \omega)}} &= \sqrt{2 \kappa} \int_{s}^t \dd t' \alpha^{(\delta \omega)}(t'|s) a_{\rm v}^\dag(t') \ket{\bm 0}, \\
   \ket{H_{t|s}^{(\delta \omega)}} &= \sqrt{2 \kappa} \int_{s}^t \dd t' \beta^{(\delta \omega)}(t'|s) a_{\rm h}^\dag(t') \ket{\bm 0}. 
\end{split}
\ee

\subsection{Visibility of a Hong-Ou-Mandel-type interference}
At this point, we know how to compute the state of the photon emitted by a single node, and we are ready to analyze the interference between photons coming from two nodes. First, note that we are only interested in events where two photons are detected at the PBSM setup. For such an event to occur (neglecting background counts), a single photon has to be emitted from both Nodes A and B, as fully captured by the non-normalized state $\mathbf{H}_t + \mathbf{V}_t$ given in Eq.~\eqref{eq: photon states}. We first model the two-photon interference by considering the cavity frequency to be fixed. We then come back to the effect of cavity jitter towards the end of this section.

\subsubsection{Two-photon state}
To fix our notation, we denote the single-photon states of Eq.~\eqref{eq: photon states 2} as $\mathbf{V}^{\rm A}_t$, $\mathbf{V}^{\rm B}_t$, $\mathbf{H}^{\rm A}_t$ and $\mathbf{V}^{\rm B}_t$ for Nodes A and B. The underlying pure states will be denoted
\be \label{eq: pure wavepackets}\begin{split}
\ket{V^{\rm A}_{t|s}} &= \sqrt{2 \kappa} \int_{s}^t \dd t' \alpha^{\rm A}(t'|s) a_{\rm v}^\dag(t') \ket{\bm 0} \\
\ket{H^{\rm A}_{t|s}} &= \sqrt{2 \kappa} \int_{s}^t \dd t' \beta^{
\rm A}(t'|s) a_{\rm h}^\dag(t') \ket{\bm 0} \\
\ket{V^{\rm B}_{t|s}} &= \sqrt{2 \kappa} \int_{s}^t \dd t' \alpha^{\rm B}(t'|s) b_{\rm v}^\dag(t') \ket{\bm 0}\\
\ket{H^{\rm B}_{t|s}} &= \sqrt{2 \kappa} \int_{s}^t \dd t' \beta^{\rm B}(t'|s) b_{\rm h}^\dag(t') \ket{\bm 0} 
\end{split}
\ee
with the natural notation for the bosonic operators $a_{\rm v}(t),a_{\rm h}(t)$ and $b_{\rm v}(t),b_{\rm h}(t)$ for Nodes A and B respectively. The scattering rates are $\text{P}_s^{\rm A}(s)$ and $\text{P}_s^{\rm B}(s)$. 
The overall density matrix $\Sigma_t$ describing the two photons (one emitted from each node) at time $t$ is thus the tensor product of the (unnormalized) states emitted from each node:
\be
\Sigma _t = (\mathbf{V}_t^{\rm A} +\mathbf{H}^{\rm A}_t)\otimes (\mathbf{V}_t^{\rm B} +\mathbf{H}^{\rm B}_t).
\ee

\subsubsection{Model of the PBSM}
\begin{figure}
    \centering
    \includegraphics[width=0.98 \columnwidth]{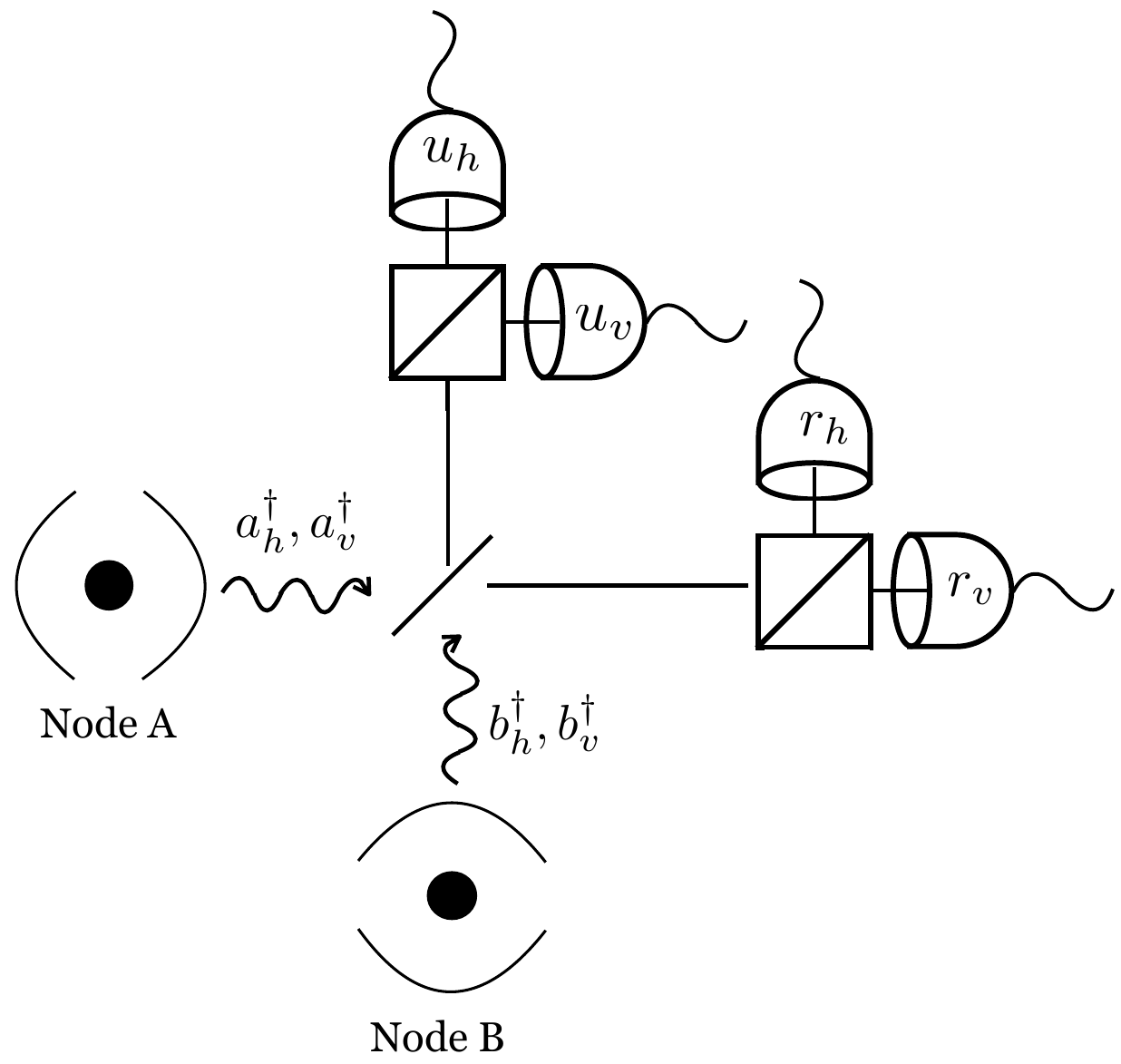}
    \caption{Representation of the detection scheme for interfering two photons from two distant nodes in a photonic Bell-state measurement (PBSM). The bosonic operators associated to the fields leaving the cavity at Node A (B) are labeled $a_h$ and $a_v$ ($b_h$ and $b_v$). The fields are combined on a nonpolarizing beamsplitter at a central station. Two detectors preceded by a polarizing beamsplitter are placed at each output of the nonpolarizing beamsplitter. The detected fields are called $u_h$, $u_v$, $r_h$ and $r_v$.}
    \label{fig:central_station}
\end{figure}

Given the two photon states, we now consider the PBSM; see Fig.~\ref{fig:central_station}. The beamsplitter output modes $u$ and $r$ are linked to the input modes $a$ and $b$ via
\be\label{eq: beamsplitter}
\binom{u}{r} =
\frac{1}{\sqrt 2}\begin{pmatrix}
1 & \ii\\
\ii & 1
\end{pmatrix} 
\binom{a}{b}
\Leftrightarrow
\binom{a}{b} =
\frac{1}{\sqrt 2}\begin{pmatrix}
1 & -\ii\\
-\ii & 1
\end{pmatrix} \binom{u}{r}.
\ee
Each output mode of the (nonpolarizing) beamsplitter consists of a polarizing beamsplitter followed by two detectors; each detector detects one of the four modes $u_h, u_v, r_h$ and $r_v$. 

Let us consider the coincidence events where two clicks occur at detectors on opposite outputs of the beamsplitter, that is, clicks at detector pairs  $\{u_h, r_h\}$, $\{u_h, r_v\}$, $\{u_v, v_h\}$ or $\{u_v, r_v\}$. We denote the rate of such coincidences for detector $u_{h(v)}$ at time $t_1$ and detector $r_{h'(v')}$ at time $t_2$ as $\text{det}_{h(v), h'(v')}(t_1,t_2)$, that is,  $\text{det}_{v,h}(t_1,t_2) \dd t^2$ corresponds to the probability to get a click at $u_v$ in the time interval $[t_1,t_1+\dd t ]$ and a click at $r_h$ in the time interval $[t_2,t_2+\dd t ]$, for example. The rate $\text{det}_{v,h}(t_1,t_2)$ corresponds to a POVM density 
\be
E_{v,h}(t_1,t_2) = \eta_{u_v} \eta_{r_h} \ketbra{ v(t_1), h(t_2)}  
\ee
with $\ket{ v(t_1), h(t_2)} = u^\dag_v(t_1) r^\dag_h(t_2) \ket{\bm 0}$,
where $\eta_{u_v}$ is the overall detection efficiency of detector $u_v$ and $\eta_{r_h}$ is the overall detection efficiency of detector $r_h$. Analogously, one defines POVM densities related to the other relevant coincidence rates $E_{h,v}(t_1,t_2)$, $E_{h,h}(t_1,t_2)$ and
$E_{v,v}(t_1,t_2)$, with 
\be\begin{split}
E_{\pi_1,\pi_2}&(t_1,t_2) =\\ &\eta_{u_{\pi_1}}\eta_{v_{\pi_2}} \, u^\dag_{\pi_1}(t_1) r^\dag_{\pi_2}(t_2)\ketbra{\bm 0} u_{\pi_1}(t_1) r_{\pi_2}(t_2),
\end{split}
\ee
which describes the event where the upper detector for polarization $\pi_1$ clicks at time $t_1$ and the right detector for polarization $\pi_2$ clicks at time $t_2$. In principle, one can also compute the probability of events where both upper detectors or both right detectors click at different times, but here we are not interested in those events.

\begin{widetext}
\subsubsection{Coincidence rates for orthogonally polarized photons}

Let us now compute the coincidence rates for two clicks from orthogonally polarized photons. We will explicitly compute the rate $\text{det}_{v,h}(t_1,t_2)$.
By the Born rule, one has
\be\begin{split}
\text{det}_{v,h}(t_1,t_2) &= \tr   \Sigma_t E_{v,h}(t_1,t_2) 
\\
& =\tr  \left(\mathbf{H}_t^{\rm A}\otimes \mathbf{H}_t^{\rm B}  + \mathbf{H}_t^{\rm A}\otimes \mathbf{V}_t^{\rm B}+ \mathbf{V}^{\rm A}_t\otimes \mathbf{H}_t^{\rm B} +\mathbf{V}^{\rm A}_t\otimes \mathbf{V}_t^{\rm B}\right)E_{v,h}(t_1,t_2) 
\\
&=
\tr (\mathbf{H}_t^{\rm A} \otimes \mathbf{V}^{\rm B}_t + \mathbf{V}_t^{\rm A} \otimes \mathbf{H}^{\rm B}_t) E_{v,h}(t_1,t_2) \\
& = \frac{\eta_{u_v} \eta_{r_{h}}}{4} \big(\tr \mathbf{H}_t^{\rm A}  a_{\rm h}^\dag(t_2)\ketbra{\bm 0}a_{\rm h}(t_2) \big) \big(\tr \mathbf{V}_t^{\rm B}  b_{\rm v}^\dag(t_1)\ketbra{\bm 0}b_{\rm v}(t_1) \big) \\
&+ \frac{\eta_{u_v} \eta_{r_h}}{4} \big(\tr \mathbf{H}_t^A  a_{\rm v}^\dag(t_1)\ketbra{\bm 0}a_{\rm v}(t_2) \big) \big(\tr \mathbf{V}_t^B  b_{\rm h}^\dag(t_1)\ketbra{\bm 0}b_{\rm h}(t_2) \big),
\end{split}
\ee
or simply
\be\label{eq: det vh}
\text{det}_{v,h}(t_1,t_2) =\frac{\eta_{u_v} \eta_{r_h}}{4}
\big(p_{\rm h}^{\rm A}(t_2) p_{\rm v}^{\rm B}(t_1)+  p_{\rm v}^{\rm A}(t_1) p_{\rm h}^{\rm B}(t_2)\big).
\ee
where $p_{\rm h}^{\rm A}(t_2)= \tr \mathbf{H}_t^{\rm A}  a_{\rm h}^\dag(t_2)\ketbra{\bm 0}a_{\rm h}(t_2)$ is the probability density that Node A emits a horizontally polarized photon at time $t_2$, and the probability densities for a vertically polarized photon and for Node B are defined equivalently. This coincidence rate can already be computed from the ion-cavity state according to Eq.~\eqref{eq:envelope} because photons of orthogonal polarization do not interfere.

Similarly, one finds
\be\label{eq: det hv}
\text{det}_{h,v}(t_2,t_1) =\frac{\eta_{u_h} \eta_{r_v}}{4}
\big(p_{\rm h}^{\rm A}(t_2) p_{\rm v}^{\rm B}(t_1)+  p_{\rm v}^{\rm A}(t_1) p_{\rm h}^{\rm B}(t_2)\big).
\ee

\subsubsection{Coincidence rates for photons with identical polarization}
It is more interesting to analyze the detection rates for two detectors sensitive to the same polarization. For example, consider $\text{det}_{h,h}(t_1,t_2)$, which is related to the projector on
\be \begin{split}
\ket{ h(t_1), h(t_2)} & =u^\dag_h(t_1) r^\dag_h(t_2) \ket{\bm 0} \\
&= \frac{1}{2}(a^\dag_{\rm h}(t_1) + \ii b^\dag_{\rm h}(t_1)) (\ii a^\dag_{\rm h}(t_2) + b^\dag_{\rm h}(t_2))\ket{\bm 0} \\
& = \frac{1}{2}(a^\dag_{\rm h}(t_1) b^\dag_{\rm h}(t_2) - a^\dag_{\rm h}(t_2)b^\dag_{\rm h}(t_1))\ket{\bm 0} + \dots
\end{split}\ee
The dots here indicate terms with two photons emitted by a single node; these terms can be ignored as $\Sigma_t$ has no support on such states. For $t_1,t_2\leq t$, the rate is thus given by 
\be\label{eq: bunch hh} \begin{split}
\text{det}_{h,h}(t_1,t_2) &= \eta_{u_h}\eta_{r_h} \tr \Sigma_t \ketbra{ h(t_1), h(t_2)} \\
& =\eta_{u_h}\eta_{r_h} \tr  \left(\mathbf{H}_t^{\rm A}\otimes \mathbf{H}_t^{\rm B}  + \mathbf{H}_t^{\rm A}\otimes \mathbf{V}_t^{\rm B}+ \mathbf{V}^{\rm A}_t\otimes \mathbf{H}_t^{\rm B} +\mathbf{V}^{\rm A}_t\otimes \mathbf{V}_t^{\rm B}\right) \ketbra{ h(t_1), h(t_2)}\\
&= \eta_{u_h}\eta_{r_h} \tr \mathbf{H}^{\rm A}_t \otimes \mathbf{H}^{\rm B}_t \ketbra{ h(t_1), h(t_2)}\\
    &= \frac{\eta_{u_h}\eta_{r_h}}{4}  \int_0^t \dd s\, \dd s' \,  \tilde{\rm P}_s^{\rm A}(s)  \tilde{\rm P}_s^{\rm B}(s')  \left| \bra{H^{\rm A}_{t|s},H^{\rm B}_{t|s'}}\big(a^\dag_{\rm h}(t_1) b^\dag_{\rm h}(t_2) - a^\dag_{\rm h}(t_2)b^\dag_{\rm h}(t_1)\big)\ket{\bm 0}\right|^2 \\
    &= \frac{\eta_{u_h}\eta_{r_h}}{4}  \int_0^t \dd s\, \dd s' \,  \tilde{\rm P}_s^{\rm A}(s)  \tilde{\rm P}_s^{\rm B}(s')  \Big| \beta^{\rm A}(t_1|s)\beta^{\rm B}(t_2|s')- \beta^{\rm A}(t_2|s)\beta^{\rm B}(t_1|s')\Big|^2.
\end{split}
\ee
Similarly,
\be\label{eq: bunch vv}
\text{det}_{v,v}(t_1,t_2) = 
\frac{\eta_{u_v}\eta_{r_v}}{4}  \int_0^t \dd s\, \dd s' \,  \tilde{\rm P}_s^{\rm A}(s)  \tilde{\rm P}_s^{\rm B}(s')  \Big| \alpha^{\rm A}(t_1|s)\alpha^{\rm B}(t_2|s')- \alpha^{\rm A}(t_2|s)\alpha^{\rm B}(t_1|s')\Big|^2.
\ee
In the integrals above, in order to use a more compact notation, we formally extend the function $\alpha(t|s)=\beta(t|s)$ to times $t<s$ by setting $\alpha(t|s)=\beta(t|s) = 0$ for $t<s$, as it is impossible for the photon to be emitted from the cavity before a scattering event to the $\ket{S,0}$ level. One can easily see from Eqs.~\eqref{eq: bunch hh} and \eqref{eq: bunch vv} that for indistinguishable pure photons, that is, $\alpha^{\rm A}(t|s)=\alpha^{\rm B}(t|s)$ and $\beta^{\rm A}(t|s)=\beta^{\rm B}(t|s)$, and no scattering, that is, $\tilde{\rm P}_s^{\rm A}(s) = \tilde{\rm P}_s^{\rm B}(s) = \delta(s)$, the photons bunch perfectly as expected at the outputs of the nonpolarizing beamsplitter, that is, $\text{det}_{h,h}(t_1,t_2)=\text{det}_{v,v}(t_1,t_2)=0$.

\end{widetext}
\subsubsection{Interference visibility}
Since we can now compute the coincidence rates at all pairs of detection times $(t_1,t_2)$ (Eqs.~\eqref{eq: bunch hh} and \eqref{eq: bunch vv}), we are also able to calculate the two-photon interference visibility. To do so, let us first define  probabilities to detect two clicks delayed by at most $T$:
\be\label{eq: Det}
\text{Det}_{\pi_1,\pi_2}(T) \equiv \int_{|t_1-t_2|\leq T} \hspace{- 5 pt} \dd t_1 \dd t_2 \,  \text{det}_{\pi_1,\pi_2}(t_1,t_2).
\ee
Then the two-photon interference visibility is by definition given by
\be
V(T) = 1- \frac{\text{Det}_{h,h}(T)+\text{Det}_{v,v}(T)}{\text{Det}_{v,h}(T)+\text{Det}_{h,v}(T)},
\ee
which one computes with the help of Eqs.~\eqref{eq: det vh}, \eqref{eq: det hv}, \eqref{eq: bunch hh}, \eqref{eq: bunch vv}, and \eqref{eq: Det}.\\

To account for the effects of the cavity jitter at Node A, we simply replace the detection probabilities above with average quantities
\be
\text{Det}_{{\pi_1,\pi_2}}(T) = \sum_{k} \text{p}(\omega_k) \text{Det}^{(\delta \omega_k)}_{{\pi_1,\pi_2}}(T),
\ee
which are obtained by averaging the detection probabilities over the possible values of $\omega_k$, as discussed previously in Sec.~\ref{sec:Eff-jitt}. 

\subsubsection{Comparison with the experimental data}
\label{se: deconstructiong}

In this section, we focus on Fig.~3b of the main text, in which the interference visibility computed with the theoretical model presented above is compared with the experimentally determined values. The figure has already been explained and discussed in the main text; our goal here is to make the connection clear between the notation used in the previous sections and the values in the plot.

The green solid line and green dashed line in Fig.~3b, which have the lowest values for visibility as a function of coincidence window, are computed with the model discussed above. The only difference between the two is the value of the cavity jitter parameter $\gamma_{clj}$ for Node A, which is given by $\gamma_{clj}= \SI{0.1}{\mega\hertz}$ for the dashed line and $\gamma_{clj}= \SI{0.06}{\mega\hertz}$ for the solid line. Both values are consistent with independently characterized experimental parameters within uncertainties.

Next, we compute the visibility expected in the absence of both laser noise ($\gamma_{ss} = 0$) and cavity jitter ($\gamma_{clj}=0$), which is plotted in orange. These are ``technical" noises that could be reduced to negligible values by realistic improvements to the setup at Node A.

Finally, the top (blue) line provides information about the role of the mismatch between pure photon wavepackets. Concretely, we consider the case $\gamma_{ss}= \gamma_{clj} = 0$ and compute the interference visibility between pure photons with the wavepackets $\ket{H_{t|0}^{\rm A(B)}}$ and $\ket{V_{t|0}^{\rm A(B)}}$ given in Eq.~\eqref{eq: pure wavepackets}, which describe the photonic states with no scattering on the $\ket{S}-\ket{P}$ transition during their evolution. The difference between the orange line and the blue line is thus solely due to the photon purity, that is, to the fact that the orange line takes into account emitted photons that are not pure due to spontaneous emission from $\ket{P}$ to $\ket{S}$. This effect can be in principle reduced by improving the coherent coupling strengths $g_1$ and $g_2$ between the ion and the cavity modes. Note that the computation of the average number of scattering events from $\ket{P}$ to $\ket{S}$ per experimental run gives $2.1$ for Node B and $5.3$ for Node A.

\bibliography{bibliography}